\documentclass[MSNbibl,nameyear,dvips]{arxstspdf}
\usepackage{mathrsfs}
\usepackage{graphicx}
\usepackage{flushend}
\usepackage{stfloats}

\volume{28}
\issue{3}
\pubyear{2013}
\firstpage{313}
\lastpage{334}
\doi{10.1214/13-STS416} 

\makeatletter

\newcommand{\eqref}[1]{(\ref{#1})}
\renewcommand{\citep}[1]{(\citeauthor{#1}, \citeyear{#1})}
\def\mathds{\mathbb}
\newcommand{\R}{\mathbb{R}}
\def\ga{\operatorname{ga}}
\def\no{\mathrm{N}}
\def\E{\mathrm{E}}
\def\d{\mathrm{d}}
\def\simind{\stackrel{\mathrm{ind}}{\sim}}
\def\simiid{\stackrel{\mathrm{i.i.d.}}{\sim}}

\newtheorem{theorem}{Theorem}
\newproclaim{remark}{Remark}
\newproclaim{definition}{Definition}

\newcommand{\Fcr}{\mathscr{F}}
\newcommand{\Dcr}{\mathscr{D}}
\newcommand{\Mcr}{\mathscr{M}}
\newcommand{\Pcr}{\mathscr{P}}
\newcommand{\X}{\mathbb{X}}
\newcommand{\Y}{\mathbb{Y}}
\newcommand{\indic}{\mathbb{I}}
\newcommand{\edr}{\mathrm{e}}
\newcommand{\ddr}{\mathrm{d}}
\newcommand{\Mc}{\mathscr{M}}
\newcommand{\tms}{\tilde{\mu}^{*}}

\newcommand{\bphi}{\bolds{\phi}}
\newcommand{\btheta}{\bolds{\theta}}
\makeatother

\begin{document}
\begin{frontmatter}

\title{Modeling with Normalized Random Measure Mixture Models}
\runtitle{NRMI Mixture Models}

\begin{aug}
\author[a]{\fnms{Ernesto} \snm{Barrios}\ead[label=e1]{ebarrios@itam.mx}},
\author[b]{\fnms{Antonio} \snm{Lijoi}}\thanksref{t1}\ead[label=e2]{lijoi@unipv.it},
\author[a]{\fnms{Luis E.} \snm{Nieto-Barajas}\corref{}\ead[label=e3]{lnieto@itam.mx}}
\and
\author[c]{\fnms{Igor} \snm{Pr\"unster}\thanksref{t1}\ead[label=e4]{igor.pruenster@unito.it}}

\runauthor{Barrios, Lijoi, Nieto-Barajas and Pr\"unster}

\affiliation{ITAM, University of Pavia, ITAM and University of Torino}

\address[a]{Ernesto Barrios is Professor and Luis E. Nieto-Barajas is Professor,
Department of Statistics, ITAM, Mexico D.F. \printead{e1,e3}.}
\address[b]{Antonio Lijoi is Professor of Probability and Mathematical Statistics, Department of Economics and
Management, University of Pavia, Italy \printead{e2}.}
\address[c]{Igor Pr\"unster is Professor of Statistics, Department of Economics and
Statistics, University of Torino, Italy \printead{e4}.}

\thankstext{t1}{Also affiliated with Collegio Carlo Alberto, Moncalieri, Italy.}

\end{aug}

%
\begin{abstract}
The Dirichlet process mixture model and more general mixtures based on
discrete random probability measures have been shown to be flexible and
accurate models for density estimation and clustering. The goal of this
paper is to illustrate the use of normalized random measures as mixing
measures in nonparametric hierarchical mixture models and point out how
possible computational issues can be successfully addressed. To this
end, we first provide a concise and accessible introduction to
normalized random measures with independent increments. Then, we
explain in detail a particular way of sampling from the posterior using
the Ferguson--Klass representation. We develop a thorough comparative
analysis for location-scale mixtures that considers a set of
alternatives for the mixture kernel and for the nonparametric
component. Simulation results indicate that normalized random measure
mixtures potentially represent a valid default choice for density
estimation problems. As a byproduct of this study an {\textsf R}
package to fit these models was produced and is available in the
Comprehensive R Archive Network (CRAN).
\end{abstract}

%

\end{frontmatter}

\section{Introduction}\label{sec1}
\label{intro}

The Dirichlet process mixture model (DPM), introduced by \citet{lo},
currently represents the most popular Bayesian nonparametric model. It
is defined as
%
\begin{equation}
\label{mix1} \tilde f(x)=\int k(x|\theta) \tilde P(\d\theta),
\end{equation}
where $k$ is a parametric kernel and $\tilde P$ is a random probability
whose distribution is the Dirichlet process prior with (finite)
parameter measure $\alpha$, in symbols $\tilde P\sim\Dcr_\alpha$.
It is
often useful to write $\alpha=a P_0$ where $P_0=\E[\tilde P]$ is a
probability measure and $a$ is in $(0,+\infty)$. In other words, the
DPM is a mixture of a kernel $k$ with mixing distribution a Dirichlet
process. See also \citet{berry} for an early contribution to DPM.

Alternatively, the DPM can also be formulated as a hierarchical model
\citep{ferg83}. In this case, $X_i,\theta_i$ for $i=1,\ldots,n$,
%
\begin{eqnarray}\label{eq:DPM}
X_i | \theta_i & \simind& k( \cdot |
\theta_i),
\nonumber
\\
\theta_i|\tilde P & \simiid& \tilde P,
\\
\tilde P & \sim& \Dcr_\alpha.
\nonumber
\end{eqnarray}
The hierarchical representation of the DPM explicitly displays features
of the model that are relevant for practical purposes. Indeed, \citet
{escobar} developed an MCMC
algorithm for simulating from the posterior distribution. This
contribution paved the way for extensive uses of the DPM, and
semiparametric variations of it, in many different applied contexts.
See \citet{mac2} and \citet{mueller} for reviews of the most remarkable achievements, both
computational and applied, in the field. The main idea behind Escobar and West's
algorithm is represented by the marginalization of the infinite
dimensional random component, namely, the Dirichlet process~$\tilde P$,
which leads to work with generalized P\'olya urn\break schemes. If the
centering measure $P_0$ is further chosen to be the conjugate prior for
kernel $k$, then one can devise a Gibbs sampler whose implementation is
straightforward. In particular, the typical setup in applications
involves a normal kernel: if the location (or location-scale) mixture
of normals 
is combined with a conjugate normal (or normal-gamma) probability
measure $P_0$, the full conditional distributions can be determined,
thus leading to a simple Gibbs sampler.

Given the importance of the DPM model, much attention has been devoted
to the development of alternative and more efficient algorithms.
According to the terminology of \citet{papa}, these can be divided into
two classes: margin\-al and conditional methods. Marginal methods, such
as the Escobar and West algorithm, integrate out the Dirichlet process
in \eqref{eq:DPM} and resort to the predictive distributions, within a
Gibbs sampler, to obtain posterior samples. In this framework an
important advance is due to \citet{mac1}: they solve the issue of
providing algorithms, which effectively tackle the case where the
kernel $k$ and $P_0$ are not a conjugate pair. On the other hand,
conditional methods work directly on \eqref{eq:DPM} and clearly have to
face the problem of sampling the trajectories of an
infinite-dimensional random element 
such as the Dirichlet process. The first\vadjust{\goodbreak} contributions along this line
are given in \citet{mulieretardella:98} and \citet{ish1} who use
truncation arguments. Exact simulations can be achieved by the
retrospective sampling technique introduced in \citet{papa} and slice
sampling schemes as in \citet{walker07}.

In this paper we focus on mixture models more general than the DPM,
namely, mixtures with mixing measure given by normalized random
measures with \mbox{independent} increments (NRMI), namely, a class of random
probability measures introduced in \citet{rlp}. Several applications of
specific members of this class, or closely related distributions, are
now present in the literature and deal with species sampling problems,
mixture models, clustering, reliability and models for dependence. See
\citet{lp} for references. Here we describe in detail a conditional
algorithm which allows one to draw posterior simulations from mixtures
based on a general NRMI. As we shall point out, it works equally well
regardless of $k$ and $P_0$ forming a conjugate pair or not and readily
yields credible intervals. Our description is a straightforward
implementation of the posterior characterization of NRMI provided in
\citet{james2} combined with the representation of an increasing
additive process given in \citet{ferg72}. The {\textsf R} package \texttt{BNPdensity},
available in the Comprehensive R Archive Network (CRAN),
implements this algorithm. For contributions containing thorough and
insightful \mbox{comparisons} of algorithms for Bayesian nonparametric mixture
models, both marginal and conditional, the reader is referred to \citet
{papa} and \citet{favaro:12}.

The \texttt{BNPdensity} package is used to carry out a comparative study
that involves a variety of data sets both real and simulated. For the
real data sets we show the impact of choosing different kernels and
compare the performance of location-scale nonparametric mixtures.
We also examine different mixing measures and show some advantages and
disadvantages fitting the data and the number of induced clusters.
Model performance is assessed by referring to conditional predictive
ordinates and to suitable numerical summaries of these values. For the
simulated examples, we rely on the relative mean integrated squared
error to measure the performance of NRMI mixtures with respect to
competing methods such as kernel density estimators, Bayesian wavelets
and finite mixtures of normals. The outcome clearly shows that NRMI\vadjust{\goodbreak}
mixtures, and in particular mixtures of stable NRMIs, potentially
represent a valid default choice for density estimation problems.

The outline of the paper is as follows. We provide in Section~\ref{sec2} an
informal review of normalized random measures and highlight their uses
for Bayesian nonparametric inference. Particular emphasis is given to
the posterior representation since it plays a key role in the
elaboration of the sampling scheme that we use; in Section~\ref{sec:algorithm} a
conditional algorithm for simulating from the posterior of NRMI
mixtures is described in great detail; Section~\ref{sec:illustration} contains a
comprehensive data analysis highlighting the potential of NRMI mixtures.

\section{Dirichlet Process and NRMIs}\label{sec2}
A deeper understanding of NRMI mixture models defined in \eqref{mix1}
is eased by an accessible introduction to the notions of completely
random measures and NRMIs. This section aims at providing a concise
review of the most relevant distributional properties of completely
random measures and NRMIs in view of their application to Bayesian
inference. These are also important for addressing the computational
issues we shall focus on in later sections.

\subsection{Exchangeability and Discrete Nonparametric Priors}\label
{sec:exch_dnp}

In order to best describe the nonparametric priors we are going to deal
with, we first recall the notion of exchangeability, its implication in
terms of Bayesian inference and some useful notation. Let $(Y_n)_{n\ge
1}$ be an (ideally) infinite sequence of observations, defined on some
probability space $(\Omega,\Fcr,P)$, with each $Y_i$ taking values in
$\Y$ (a complete and separable metric space endowed with its Borel
$\sigma$-algebra). While in a frequentist setting one typically
assumes that the $Y_i$'s are independent and identically distributed
(i.i.d.) with some fixed and unknown distribution, in a Bayesian approach
the independence is typically replaced by a weaker assumption of
conditional independence, given a random probability distribution on
$\Y
$, which corresponds to assuming \emph{exchangeable} data.
Formally, this corresponds to an invariance condition according to
which, for any $n\ge1$ and any permutation $\pi$ of the indices
$1,\ldots,n$, the probability distribution of $(Y_1,\ldots,Y_n)$
coincides with the distribution of $(Y_{\pi(1)},\ldots,Y_{\pi(n)})$.
Then, the celebrated de Finetti representation theorem states that the
sequence $(Y_n)_{n\ge1}$ is exchangeable if and only if its
distribution can be represented as a mixture of sequences of i.i.d. random
variables. In other terms, $(Y_n)_{n\ge1}$ is exchangeable if and only
if there exists a probability distribution $Q$ on the space of
probability measures on $\Y$, say, $\Pcr_\Y$, such that
%
\begin{eqnarray}\label{eq:nonpmic}
Y_i | \tilde P &\simiid& \tilde P,\quad i=1,\ldots,n,
\nonumber
\\[-8pt]
\\[-8pt]
\nonumber
\tilde P &\sim& Q
\end{eqnarray}
for any $n\ge1$. Hence, $\tilde P$ is a random probability measure on
$\Y$, namely, a random element on $(\Omega,\Fcr,P)$ taking values in
$\Pcr_\Y$ (endowed with the topology of weak convergence). The
probability distribution $Q$ of $\tilde P$ is also termed \textit
{de Finetti measure} and represents the prior distribution in a
Bayesian setup. Whenever $Q$ degenerates on a finite-dimensional
subspace of $\Pcr_\Y$, the inferential problem is usually called
\textit
{parametric}. On the other hand, when the support of $Q$ is infinite
dimensional, then this is typically referred to as a \textit
{nonparametric} inferential problem. It is generally agreed that having
a large topological support is a desirable property for a nonparametric
prior (see, e.g., \cite*{Feg74}).

In the context of nonparametric mixture models, which identify the main
focus of the paper, a key role is played by \emph{discrete
nonparametric priors} $Q$, that is, priors which select discrete
distributions with probability~$1$. Clearly, any random probability
measure $\tilde P$ associated to a discrete prior $Q$ can be
represented as
%
\begin{equation}
\label{eq:discrete} \tilde P=\sum_{j \geq1} \tilde
p_j \delta_{Z_j},
\end{equation}
where $(\tilde p_j)_{j\ge1}$ is a sequence of nonnegative random
variables such that $\sum_{j\ge1}\tilde p_j=1$, almost surely,\break
$(Z_j)_{j\ge1}$ is a sequence of random variables taking values in $\Y
$ and $\delta_Z$ is the Dirac measure.

As far as the observables $Y_i$'s are concerned, the discrete nature of
$\tilde P$ in \eqref{eq:discrete} implies that any sample $Y_1, \ldots,
Y_n$ in \eqref{eq:nonpmic} will feature ties with positive probability
and, therefore, display $r \leq n$ distinct observations $Y_1^*, \ldots
, Y_{r}^*$ with respective frequencies $n_1, \ldots, n_r$ such that
$\sum_{i=1}^r n_i=n$. Such a grouping lies at the heart of Bayesian nonparametric
procedures for clustering purposes. Henceforth, $R_n$ will denote the
random variable identifying the number of distinct values appearing in
the sample $Y_1,\ldots,Y_n$.

The simplest and most familiar illustration one can think of is the
Dirichlet process prior introduced by \citet{Feg73}, which represents
the cornerstone of Bayesian Nonparametrics. Its original definition was
given in terms of a consistent family of finite-dimensional
distributions that coincide with multivariate Dirichlet distributions.
To make this explicit, introduce the $(d-1)$-variate Dirichlet\vadjust{\goodbreak}
probability density function on the $(d-1)$-dimensional unit simplex
\begin{eqnarray*}
h(\mathbf{p};\mathbf{c})&=&\frac{\Gamma(\sum_{i=1}^d c_i)}{\prod_{i=1}^d
\Gamma
(c_i)} p_1^{c_1-1} \cdots
p_{d-1}^{c_{d-1}-1}\\
&&{}\cdot \Biggl(1-\sum_{i=1}^{d-1}p_i
\Biggr)^{c_d-1}, 
\end{eqnarray*}
where $\mathbf{c}\!=\!(c_1,\ldots,c_d)\!\in\!(0,\infty)^d$ and $\mathbf{p}\!=\!(p_1,\ldots, p_{d-1})$.

\begin{definition}[(Ferguson, \citeyear{Feg73})]\label{def:dirichl_fd}
Let $\alpha$ be some finite and nonnull measure on $\Y$ such that
$\alpha(\Y)=a$. Suppose the random probability measure $\tilde P$ has
distribution $Q$ such that, for any choice of a (measurable) partition
$\{A_1,\ldots,A_d\}$ of $\Y$ and for any $d\ge1$, one has
%
\begin{eqnarray}
\label{eq:fd_dirichlet} &&Q \bigl( \bigl\{P\dvtx \bigl(P(A_1),
\ldots,P(A_d) \bigr)\in B \bigr\} \bigr)
\nonumber
\\[-8pt]
\\[-8pt]
\nonumber
&&\quad= \int_B
h(\mathbf{p};\bolds{\alpha}) \,\ddr p_1 \cdots \,\ddr p_{d-1},
\end{eqnarray}
where $\bolds{\alpha}=(\alpha(A_1),\ldots,\alpha(A_d))$. Then $\tilde P$
is termed a \textup{Dirichlet process} with base measure $\alpha$.
\end{definition}

Note that $\alpha/a=:P_0$ defines a probability measure on $\Y$ and it
coincides with\vspace*{1pt} the expected value of a Dirichlet process, that is,
$P_0=\E[\tilde P]$, for this reason is often referred to as the prior
guess at the shape of~$\tilde P$. Henceforth, we shall denote more
conveniently the base measure $\alpha$ of $\tilde P$ as $a P_0$. Also
note that the Dirichlet process has large support and it, thus, shares
one of the properties that makes the use of nonparametric priors attractive.
Indeed, if the support of $P_0$ coincides with $\Y$, then the support
of the Dirichlet process prior (in the weak convergence topology)
coincides with the whole space $\Pcr_\Y$. In other words, the Dirichlet
process prior assigns positive probability to any (weak) neighborhood
of any given probability measure in $\Pcr_\Y$, thus making it a
flexible model for Bayesian nonparametric inference.

As shown in \citet{black73}, the Dirichlet process selects discrete
distributions on $\Y$ with probability 1 and, hence, admits a
representation of the form \eqref{eq:discrete}. An explicit
construction of the $\tilde p_j$'s in \eqref{eq:discrete} leading to
the Dirichlet process has been provided by \citet{Set94} who relied on a
stick-breaking procedure. This arises when the sequence of random
probability masses $(\tilde p_j)_{j\ge1}$ is defined as
%
\begin{eqnarray}
\label{eq:stick_weights} \tilde p_1&=&V_1,
\nonumber
\\[-8pt]
\\[-8pt]
\nonumber
\tilde p_j&=&V_j \prod_{i=1}^{j-1}(1-V_i),\quad
j=2,3,\ldots,
\end{eqnarray}
with the $V_i$'s being i.i.d. and beta distributed with parameter $(1,a)$,
and when
the locations $(Z_j)_{j\ge1}$ are i.i.d. from $P_0$. Under these
assumptions \eqref{eq:discrete} yields a random probability measure
that coincides, in distribution, with a Dirichlet process with base
measure $a  P_0$.

A nice and well-known feature about the Dirichlet process is its
conjugacy. Indeed, if $\tilde P$ in \eqref{eq:nonpmic} is a Dirichlet
process with base measure $a  P_0$, then the posterior distribution of
$\tilde P$, given the data $Y_1,\ldots,Y_n$, still coincides with the
law of a Dirichlet process with parameter measure $(a+n) P_n$ where
$P_n=a  P_0/\break(a+n)+\sum_{i=1}^n\delta_{Y_i} /(a+n)$, where $\delta_y$
denotes a point mass at $y\in\Y$. On the basis of this result, one
easily determines the predictive distributions associated to the
Dirichlet process and for any $A$ in $\Y$, one has
%
\begin{eqnarray}
\label{eq:pred_dir} &&P [Y_{n+1}\in A | Y_1,
\ldots, Y_n ]
\nonumber
\\[-8pt]
\\[-8pt]
\nonumber
&&\quad=\frac{a}{a+n} P_0(A)+\frac{n}{a+n}
\frac{1}{n} \sum_{j=1}^r
n_j \delta_{Y_j^*}(A),
\end{eqnarray}
where, again, the $Y_j^*$'s with frequency $n_j$ denote the $r\leq n$
distinct observations within the sample. Hence, the predictive
distribution appears as a convex linear combination of the prior guess
at the shape of $\tilde P$ and of the empirical distribution.

From \eqref{eq:discrete} it is apparent that a decisive issue when
defining a discrete nonparametric prior is the determination of the
probability masses $\tilde p_j$'s, while at the same time preserving a
certain degree of mathematical tractability. This is in general quite a
challenging task. For instance, the stick-breaking procedure is useful
to construct a wide range of discrete nonparametric priors as shown in
\citet{ish1}. However, only for a few of them is it possible to
establish relevant distributional properties such as, for example, the
posterior or predictive structures. See \citet{Fav12} for a discussion
on this issue. Also, as extensively discussed in \citet{lp}, a~key tool
for defining tractable discrete nonparametric priors \eqref
{eq:discrete} is given by completely random measures, a~concept
introduced in \citet{king67}. Since it is essential for the construction
of the class of NRMIs considered in the paper, in the following section
we concisely recall the basics and refer the interested reader to \citet
{king93} for an exhaustive account.

\subsection{CRM and NRMI}\label{sec:cnrmi}

Denote first by $\Mc_\Y$ the space of boundedly finite measures on
$\Y
$, this meaning that for any $\mu$ in $\Mc_\Y$ and any bounded set $A$
in $\Y$ one has $\mu(A)<\infty$. Moreover, $\Mcr_\Y$ can be endowed
with a suitable topology that allows one to define the associated Borel
$\sigma$-algebra. See \citet{DaleyVereJones:1988} for technical details.

\begin{definition}
A random element $\tilde\mu$, defined on $(\Omega,\Fcr,P)$ and
taking values in $\Mc_\Y$, is called a \emph{completely random
measure} (CRM) if, for any $n\ge 1$ and $A_1,\ldots,A_n$ in $\Y$, with $A_i\cap
A_j=\varnothing$ for any $i\ne j$, the random variables
$\tilde\mu(A_1),\ldots,\tilde\mu(A_n)$ are mutually independent.
\end{definition}

Hence, a CRM is simply a random measure, which gives rise to
independent random variables when evaluated over disjoint sets. In
addition, it is well known that if $\tilde\mu$ is a CRM on $\Y$, then
\[
\tilde\mu=\sum_{i\ge1}J_i
\delta_{Z_i}+\sum_{i=1}^M
V_i \delta_{z_i},
\]
where $(J_i)_{i\ge1}$, $(V_i)_{i\ge1}$ and $(Z_i)_{i\ge1}$ are
independent sequences of random variables and the jump points $\{
z_1,\ldots,z_M\}$ are fixed, with $M\in\{0,1,\ldots\}\cup\{\infty\}$.
If $M=0$, then $\tilde\mu$ has no fixed jumps and the Laplace transform
of $\tilde\mu(A)$, for any $A$ in $\Y$, admits the following representation:
%
\begin{eqnarray}
\label{eq:lapl_transform} &&\E \bigl[\edr^{-\lambda\tilde\mu(A)} \bigr]
\nonumber
\\[-8pt]
\\[-8pt]
\nonumber
&&\quad=\exp
\biggl\{- \int_{\R^+\times A} \bigl[1-\edr^{- \lambda v } \bigr] \nu(
\ddr v,\ddr y) \biggr\}
\end{eqnarray}
for any $\lambda>0$, with $\nu$ being a measure on $\R^+\times\Y$
such that
%
\begin{equation}
\label{eq:levy_intensity} \int_B\int
_{\R^+}\min\{v,1\} \nu(\ddr v,\ddr y)<\infty
\end{equation}
for any bounded $B$ in $\Y$. The measure $\nu$ is referred to as the
\emph{L\'evy intensity} of $\tilde\mu$ and, by virtue of \eqref
{eq:lapl_transform}, it characterizes the CRM $\tilde\mu$. 
This is extremely useful from an operational point of view since a
single measure encodes all the information about the distribution of
the jumps $(J_i)_{i\ge1}$ and locations $(Z_i)_{i\ge1}$ of $\tilde
\mu
$. The measure $\nu$ will be conveniently rewritten as
%
\begin{equation}
\label{pim} \nu(\ddr v,\ddr y)=\rho(\ddr v| y) \alpha(\ddr y),
\end{equation}
where $\rho$ is a transition kernel on $\R^+ \times\Y$ controlling
the jump intensity and $\alpha$ is a measure on $\Y$ determining the
locations of the jumps. Two popular examples are \textit{gamma} and
\textit{stable} processes. The former corresponds to the specification
$\rho(\ddr v|y)=\edr^{-v} \,\ddr v/v$, whereas the latter arises when
$\rho(\ddr v|y)=\gamma  v^{-1-\gamma} \,\ddr v/\Gamma(1-\gamma)$, for
some $\gamma\in(0,1)$. Note that if $\tilde\mu$ is a gamma CRM, then,\vadjust{\goodbreak}
for any $A$, $\tilde\mu(A)$ is gamma distributed with shape parameter
$\alpha(A)$ and scale $1$. On the other hand, if $\tilde\mu$ is a
stable CRM, then $\tilde\mu(A)$ has a positive stable distribution.

Since $\tilde\mu$ is a discrete random measure almost surely, one can
then easily guess that discrete random probability measures \eqref
{eq:discrete} can be obtained by suitably transforming a CRM. 
The most obvious transformation is ``normalization,'' which yields
NRMIs. As a preliminary remark, it should be noted that
``normalization'' is possible when the denominator $\tilde\mu(\Y)$ is
positive and finite (almost surely). Such a requirement can be
expressed in terms of the L\'evy intensity, in particular, $\alpha$
being a finite measure
and $\int_{\R^+}\rho(\d v|y)=\infty$ for any $y \in\Y$ are simple
sufficient conditions for the normalization to be well defined. The
latter condition essentially requires the CRM to jump infinitely often
on any bounded set and is sometimes referred to as \emph{infinite
activity}. See \citet{rlp} and \citet{james2} for necessary and
sufficient conditions. One can now provide the definition of a NRMI.

\begin{definition}
Let $\tilde\mu$ be a CRM with L\'evy intensity \eqref{pim} such that
$0<\tilde\mu(\Y) < \infty$ almost surely. Then, the random probability
measure
%
\begin{equation}
\label{eq:NRMI} \tilde P =\frac{\tilde\mu}{\tilde\mu(\Y)}
\end{equation}
is named a normalized random measure with independent increments (NRMI).
\end{definition}

It is apparent that a NRMI is uniquely identified by the L\'evy
intensity $\nu$ of the underlying CRM. If $\rho(\d v|y)$ in \eqref{pim}
does not depend on $y$, which means that the distribution of the jumps
of $\tilde\mu$ are independent of their locations, then the CRM
$\tilde\mu$ and the corresponding NRMI \eqref{eq:NRMI} are called
\textit{homogeneous}. Otherwise they are termed \textit
{nonhomogeneous}. Moreover, it is worth pointing out that all NRMI
priors share a support property analogous to the one recalled for the
Dirichlet process prior. Specifically, if the support of the base
measure coincides with $\Y$, then the corresponding NRMI has full weak
support~$\mathscr{P}_{\Y}$.

Note that the Dirichlet process can be defined as an NRMI: indeed, it
coincides, in distribution, with a normalized gamma CRM as shown in
\citet{Feg73}. If $\nu(\ddr v,\ddr y)=\edr^{-v} v^{-1} \,\ddr v\, a
P_0(\ddr y)$, then \eqref{eq:NRMI} yields a Dirichlet process with base
measure $a  P_0$.
Another early use of \eqref{eq:NRMI} can be found in \citet{kingman},
where the NRMI obtained by normalizing a stable CRM is introduced. The
resulting random probability measure will be denoted as N-stable.\vadjust{\goodbreak}

In the sequel particular attention will be devoted to generalized gamma
NRMIs (Lijoi, Mena and\break Pr{\"u}nster, \citeyear{lijoi07}) since they are analytically tractable and include
many well-known priors as special cases. This class of NRMIs is
obtained by normalizing generalized gamma CRMs that were introduced in
\citet{brix} and are characterized by a L\'evy intensity of the form
%
\begin{equation}
\label{eq:ngg} \rho(\d v) \alpha( \d x) =\frac{\edr^{-\kappa v}}{\Gamma(1-\gamma)  v^{1+\gamma}}\,\d v\, a
P_0(\d x),
\end{equation}
whose parameters $\kappa\geq0$ and $\gamma\in[0,1)$ are such that at
least one of them is strictly positive
and with base measure $\alpha=a P_0$, where $a \in(0, \infty)$ and
$P_0$ is a probability distribution on $\Y$. The corresponding
generalized gamma NRMI will be denoted as $\tilde P \sim\operatorname
{NGG}(a,\kappa,\gamma; P_0)$. Within this class of priors one finds the
following special cases: (i) the Dirichlet process which is a
$\operatorname
{NGG}(a, 1, 0; P_0)$ process; (ii)~the normalized inverse Gaussian
(N-IG) process (Lijoi, Mena and Pr{\"u}nster, \citeyear{lijoi05}), which corresponds to a $\operatorname{NGG}(1,
\kappa,1/2; P_0)$ process; (iii) the \mbox{N-stable} process \citep{kingman}
which arises as $\operatorname{NGG}(1, 0 ,\gamma; P_0)$. As a side
remark, we
observe 
that either $\kappa$ or $a$ can be fixed according to one's
convenience. Loosely speaking, this is due to the fact that the
normalization operation implies the loss of ``one degree of freedom''
as a reference to the Dirichlet process might clarify. For example, we
mentioned that the Dirichlet case arises when $\kappa$ is set equal to~$1$, but this choice is only due to convenience. Indeed, a~Dirichlet
process is obtained, as long as $\gamma=0$, whatever the value $\kappa$
takes on. See \citet{pitman} and \citet{lijoi07} for detailed
explanations. For our purposes it is worth sticking to the redundant
parameterization since it allows us to recover immediately all three
specific cases listed above, which would be cumbersome with the
alternative parameterization usually adopted, that is, $\operatorname{NGG}(1,
\beta, \gamma; P_0)$ with $\beta=\kappa^\gamma/\gamma$.
The role of these parameters is best understood by looking at the
induced (prior) distribution of the number of distinct values $R_n$ in
an sample $Y_1,\ldots,Y_n$. Indeed, one has that $\kappa$ (or,
equivalently, $a$) affects the location: a larger $\kappa$ (or $a$)
shifts the distribution of $R_n$ to the right, implying a larger
expected number of distinct values. In contrast, $\gamma$ allows to
tune the flatness of the distribution of $R_n$: the bigger $\gamma$,
the flatter is the distribution of $R_n$ so that a large value of
$\gamma$ corresponds to a less informative prior for the number of
distinct values in $Y_1,\ldots,Y_n$. This also explains why the
Dirichlet process, which corresponds to $\gamma=0$, yields the most
highly-peaked distribution for $R_n$. See also \citet{lijoi07} for a
graphical display of these behaviors.

Also, variations of NRMI have already appeared in the literature. In
\citet{npw} weighted versions of NRMIs are considered. To be more
specific, letting $h$ be some nonnegative function defined on $\Y$, a
normalized weight\-ed CRM is obtained, for any $B$ in $\Y$, as
\[
\tilde P(B)=\frac{\int_{B} h(y) \tilde\mu(\d y)}{\int_{\Y} h(y)
\tilde
\mu(\d y)}.
\]
The function $h$ can be seen as a perturbation of the CRM and in \citet
{nbp2} the sensitivity of posterior inference with respect to (w.r.t.)
$h$ is examined. Another related class is represented by
Poisson--Kingman models \citep{pitman}, where one essentially
conditions on $\tilde\mu(\Y)$ and then mixes with respect to some
probability measure on~$\R^+$.

\begin{remark}\label{rem:cdf}
If $\Y=\mathds{R}^m$, one can also consider the c\`adl\`ag random
distribution function induced by $\tilde\mu$, namely, $\tilde M:=\{
\tilde M(s)=\tilde\mu((-\infty,s_1] \times\cdots \times(-\infty,
s_m])\dvtx\break s=(s_1, \ldots, s_m) \in\R^m\}$, known in the literature as the
\textit
{increasing additive process} or \emph{independent increment process}.
See \citet{sato} for details. One can then associate to the NRMI random
probability measure in \eqref{eq:NRMI} the corresponding NRMI random
cumulative distribution function
%
\begin{equation}
\label{eq:P1bis} \tilde F(s)=\frac{\tilde M(s)}{T}\quad \mbox{for any } s \in
\R^m,
\end{equation}
where $T:=\lim_{s\to\infty} \tilde M(s)$ and the limit is meant as
componentwise. The original definition of NRMI in \citet{rlp} was given
in terms of increasing additive processes. The definition on more
abstract spaces adopted here, and used also, for example, in \citet
{james2}, allows us to bypass some tedious technicalities. Nonetheless,
we preserve the term NRMI, although on abstract spaces one should refer
to normalized CRM rather than to ``increments.''
\end{remark}

\begin{remark}$\!\!\!$
Although the previous examples deal with homogeneous CRMs and NRMIs,
nonhomogeneous CRMs are also very useful for the construction of
nonparametric priors. This is apparent in contributions to Bayesian
nonparametric inference for survival analysis. See \citet{lp}. Hence,
given the
importance of nonhomogeneous structures in some other contexts, it
seems worth including these in our treatment.
\end{remark}

\subsection{Posterior Distribution of a NRMI}

The posterior distribution associated to an exchangeable model as in
\eqref{eq:nonpmic} is a preliminary step for attaining Bayesian
inferential results of interest and, therefore, represents an object of
primary importance. In the case of NRMIs, the determination of the
posterior distribution is a challenging task since one cannot rely
directly on Bayes' theorem (the model is not dominated) and, with the
exception of the Dirichlet process, \mbox{NRMIs} are not conjugate as shown in
\citet{james1}. Nonetheless, a posterior characterization has been
established in \citet{james2} and it turns out that, even though NRMIs
are not conjugate, they still enjoy a sort of ``conditional
conjugacy.'' This means that, conditionally on a suitable latent random
variable, the posterior distribution of a NRMI coincides with the
distribution of a NRMI having fixed points of discontinuity located at
the observations. Such a simple structure suggests that when working
with a general NRMI, instead of the Dirichlet process, one faces only
one additional layer of difficulty represented by the marginalization
with respect to the conditioning latent variable.

Before stating the main result we recall that, due to the discreteness
of NRMIs, ties will appear with positive probability in 
${\mathbf Y}=(Y_1, \ldots, Y_n)$ and, therefore, the sample information can
be encoded by the $R_n=r$ distinct observations $(Y_1^*,\ldots,Y_r^*)$
with frequencies $(n_1, \ldots, n_r)$ such that $\sum_{j=1}^r n_j=n$.
Moreover, introduce the nonnegative random variable $U$ such that the
distribution of $[U|\mathbf{Y}]$ has density, w.r.t. the Lebesgue measure,
given by
%
\begin{equation}
\label{eq:U}\quad f_{U|{\mathbf Y}}(u)\propto u^{n-1}\exp \bigl\{-\psi(u)
\bigr\}\prod_{j=1}^r\tau _{n_j}
\bigl(u|Y_j^* \bigr),
\end{equation}
where $\tau_{n_j}(u|Y_j^*)=\int_0^\infty v^{n_j}\mathrm{e}^{-u v}\rho(\d
v|Y_j^*)$ and $\psi$ is the Laplace exponent of $\tilde\mu$ as in
\eqref{eq:lapl_transform}. Finally, in the following we assume the
probability measure $P_0$ defining the base measure of a NRMI to be nonatomic.

\begin{theorem}[\citep{james2}]\label{james2}\break
Let $(Y_n)_{n \geq1}$ be as in \eqref{eq:nonpmic} where $\tilde P$
is a NRMI defined in \eqref{eq:NRMI} with L\'evy intensity as in
\eqref
{pim}. Then the posterior distribution of the unnormalized CRM $\tilde
\mu$, given a sample ${\mathbf Y}$, is a mixture of the distribution of
$[\tilde\mu|U,\mathbf{Y}]$ with respect to the distribution of $[U|\mathbf
{Y}]$. The latter is identified by \eqref{eq:U},\vadjust{\goodbreak} whereas $[\tilde\mu
|U,\mathbf{Y}]$ is equal in distribution to a CRM with fixed points of
discontinuity at the distinct observations $Y_j^*$,
%
\begin{equation}
\label{eq:post_CRM} \tilde\mu^*+\sum_{j=1}^r
J_j^*\delta_{Y_j^*}
\end{equation}
such that:
\begin{longlist}[(a)]
\item[(a)] $\tms$ is a CRM characterized by the L\'evy intensity
%
\begin{equation}
\label{eq:post_levy} \nu^*(\d v, \d y)=\mathrm{e}^{-u v}
\rho(\d v|y)\alpha(\d y);
\end{equation}
\item[(b)]
the jump height $J_j^*$ corresponding to $Y_j^*$ has density, w.r.t.
the Lebesgue measure, given by
%
\begin{equation}
\label{eq:post_jump} f_{j}^*(v)\propto
v^{n_j}\mathrm{e}^{-u v}\rho \bigl(\d v|Y_j^* \bigr).
\end{equation}
\item[(c)] $\tms$ and $J_j^*$, $j=1,\ldots,r$, are
independent.
\end{longlist}

Moreover, the posterior distribution of the NRMI $\tilde P$,
conditional on $U$, is given by
%
\begin{equation}
\label{eq:post.ncrm}\quad [\tilde P | U, \mathbf{Y}] \stackrel{d} {=} w
\frac{\tilde\mu^*}{\tilde\mu^*(\X)}+(1-w) \frac{\sum_{i=1}^r
J_i^*
\delta_{Y_i^*}}{\sum_{l=1}^r J_l^*},
\end{equation}
where $w=\tilde\mu^*(\X)/(\tilde\mu^*(\X)+\sum_{l=1}^r J_l^{*})$.
\end{theorem}

In order to simplify the notation, in the statement we have omitted
explicit reference to the dependence on $[U|\mathbf{Y}]$ of both $\tilde
\mu
^*$ and $\{J_i^*\dvtx i=1,\ldots,r\}$. However, such a dependence is
apparent from \eqref{eq:post_levy} and~\eqref{eq:post_jump}. 
From Theorem~\ref{james2} follows is apparent that the only quantity needed for
deriving explicit expressions for particular cases of NRMI is the L\'
evy intensity \eqref{pim}. For instance, in the case of normalized
generalized gamma NRMI, $\operatorname{NGG}(a,\kappa,\gamma; P_0)$
one has that
the unnormalized posterior CRM $\tms$ in \eqref{eq:post_CRM} is
characterized by a L\'evy intensity of the form
%
\begin{equation}
\label{eq:post_levy_ngg} \nu^*(\d v,\d y)=
\frac{\edr^{-(\kappa+u)v}}{\Gamma(1-\gamma)
v^{1+\gamma}}\,\d v\, a P_0(\d y).
\end{equation}
Moreover, the distribution of the jumps \eqref{eq:post_jump}
corresponding to the fixed points of discontinuity $Y_i^*$'s in \eqref
{eq:post_CRM} reduce to a gamma distribution with density
%
\begin{equation}
\label{postfj} f_{j}^*(v)=\frac{(\kappa+u)^{n_j-\gamma}}{\Gamma(n_j-\gamma)} v^{n_j-\gamma-1}
\mathrm{e}^{-(\kappa+u)v}. 
\end{equation}
Finally, the conditional distribution of the latent variable $U$ given
${\mathbf Y}$ \eqref{eq:U} is given by
%
\begin{eqnarray}
\label{condu} &&f_{U|{\mathbf Y}}(u)
\nonumber
\\[-8pt]
\\[-8pt]
\nonumber
&&\quad\propto u^{n-1}(u+
\kappa)^{r\gamma-n}\exp \biggl\{ -\frac
{a}{\gamma}(u+\kappa)^{\gamma}
\biggr\}
\end{eqnarray}
for $u>0$.
The availability of this posterior characterization makes it then
possible to determine several\vadjust{\goodbreak} important quantities such as the
predictive distributions and the induced partition distribution. See
\citet{james2} for general NRMI and \citet{lijoi07} for the subclass of
generalized gamma NRMI.

\subsection{NRMI Mixture Models}\label{sec:mixture}

Discrete nonparametric priors are particularly effective when used for
modelling latent variables\break within hierarchical mixtures. 
The most popular of these models is the DPM due to \citet{lo} and
displayed in \eqref{eq:DPM}. Its most natural generalization
corresponds to allowing any NRMI to act as a nonparametric mixing
measure. In view of the result on the posterior characterization of
NRMIs, such a program is also feasible from a practical perspective.

We start by describing the NRMIs mixture model in some detail. First,
let us introduce a change in the notation. In order to highlight that
the law of a \mbox{NRMIs} acts as the de Finetti measure at a latent level, we
denote the elements of the exchangeable sequence by $\theta_i$ instead
of $Y_i$, for $i=1,2, \ldots.$ Then, consider a NRMI $\tilde P$ and
convolute it with a suitable density kernel $k(  \cdot  | \theta)$,
thus obtaining the random mixture density $\tilde f(x)=\int_{\Theta}
k(x|\theta) \tilde P(\ddr\theta)$. This can equivalently be written in
a hierarchical form as
%
\begin{eqnarray}\label{eq:mixt}
X_i | \theta_i & \simind& k( \cdot |
\theta_i),\quad i=1,\ldots,n,
\nonumber
\\
\theta_i|\tilde P & \simiid& \tilde P,\quad i=1,\ldots,n,
\\
\tilde P & \sim& \operatorname{NRMI}.
\nonumber
\end{eqnarray}
In the sequel, we take kernels defined on $\X\subseteq\R$ and NRMIs
defined on $\Y=\Theta\subseteq\R^m$. Consequently, instead of
describing the results in terms of the random measures $\tilde\mu$ and
$\tilde P$, we will work with corresponding distribution functions
$\tilde M$ and $\tilde F$, respectively, for the sake of simplicity in
the presentation (see Remark~\ref{rem:cdf}). It is worth noting that
the derivations presented here carry over to general spaces in a
straightforward way.

As for the base measure of the NRMI $P_0$ on $\Theta$, we denote its
density (w.r.t. the Lebesgue measure) by $f_0$. When $P_0$ depends on a
further hyperparameter $\phi$, we will use the symbol $f_0(  \cdot |
\phi)$. The case $m=2$ typically corresponds to the specification of a
nonparametric model for the 
location and scale parameters of the mixture, that is, $\theta=(\mu
,\sigma)$. This will be used to illustrate the algorithm in Section
\ref
{sec:illustration}, where we apply our proposed modeling to simulated
and real data sets. In order to distinguish the hyperparameters for
location and scale, we will use the notation $f_0(\mu,\sigma|\phi
)=f_{0}^1(\mu|\sigma, \varphi) f_{0}^2(\sigma| \varsigma)$. In
applications a priori independence between $\mu$ and $\sigma$ is
commonly assumed.

The most popular uses of mixtures of discrete random probability
measures, such as the one displayed in \eqref{eq:mixt}, relate to
density estimation and data clustering. The former can be addressed by
evaluating
%
\begin{equation}
\hat{f}_n(x)=\E \bigl(\tilde f(x) | X_1,
\ldots,X_n \bigr) \label{eq:density_est}
\end{equation}
for any $x$ in $\X$.
As for the latter, if $R_n$ is the number of distinct latent values
$\theta_1^*,\ldots,\theta_{R_n}^*$ out of a sample of size~$n$, one can
deduce a partition of the observations such that any two $X_i$ and
$X_j$ belong to the same cluster if the corresponding latent variables
$\theta_i$ and $\theta_j$ coincide. Then, it is interesting to
determine an estimate $\hat{R}_n$ of the number of clusters into which
the data are grouped. In the examples we will illustrate $\hat{R}_n$ is
set equal to the mode of $R_n|\mathbf{X}$, with $\mathbf{X}:=(X_1,\ldots,X_n)$
representing the observed sample. Both estimation problems can be faced
by relying on the simulation algorithm that will be detailed in the
next section.

\section{Posterior Simulation of NRMI Mixtures}\label{sec:algorithm}

Our main aim is to provide a general algorithm to draw posterior
inferences with the mixture model~\eqref{eq:mixt}, for any choice of
the mixing NRMI and of the kernel. A further byproduct of our algorithm
is the possibility of determining credible intervals.
The main block of the conditional algorithm presented in this section
is the posterior representation provided in Theorem~\ref{james2}. In fact, in
order to sample from the posterior distribution of the random mixture
model~\eqref{eq:mixt}, given a sample $X_1,\ldots,X_n$, a
characterization of the posterior distribution of the mixing measure at
the higher stage of the hierarchy is needed. We rely on the posterior
representation, conditional on the unobservable variables ${\btheta
}:=(\theta_1,\ldots,\theta_n)$, of the unnormalized process~$\tilde M$,
since the normalization can be carried out within the algorithm.

For the implementation of a Gibbs sampling scheme we use the
distributions of
%
\begin{equation}
\label{cdist} [\tilde M|{\mathbf X},{\btheta}] \quad\mbox{and}\quad [{\btheta}|{\mathbf
X},\tilde M].
\end{equation}
For illustration we shall detail the algorithm when $\tilde P \sim
\operatorname
{NGG}(a,\kappa,\gamma; P_0)$ and provide explicit expressions for each
of the distributions in \eqref{cdist}. Nonetheless, as already
recalled, the algorithm can be implemented for any NRMI: one just needs
to plug in the corresponding L\'evy intensity.

Due to conditional independence properties, the conditional
distribution of $\tilde M$, given ${\mathbf X}$\vadjust{\goodbreak} and ${\btheta}$, does not
depend on ${\mathbf X}$, that is, $[\tilde M|{\mathbf X},{\btheta}]=[\tilde
M|{\btheta}]$. Now, by Theorem~\ref{james2}, the posterior distribution function
$[\tilde M|{\btheta}]$ is characterized as a mixture in terms of a
latent variable $U$, that is, through $[\tilde M|U,{\btheta}]$ and
$[U|{\btheta}]$. Specifically, the conditional distribution of $\tilde
M$, given $U$ and ${\btheta}$, is another CRM with fixed points of
discontinuity at the distinct $\theta_i$'s, namely, $\{\theta
_1^*,\ldots
,\theta_r^*\}$, given by
%
\begin{equation}
\label{conda} \tilde M^*_+(s) :=\tilde M^*(s)+\sum_{j=1}^r
J_j^* \indic_{(-\infty
,s]} \bigl(\theta_j^* \bigr),
\end{equation}
where $(-\infty,s]=\{y\in\R^m\dvtx y_i\leq s_i,   i=1,\ldots, m\}$
and $\indic_{A}$ denotes the indicator
function of a set $A$.
Recall that in the $\operatorname{NGG}(a,\kappa,\gamma; P_0)$ case,
$\tilde
M^*$ has L\'evy intensity as in \eqref{eq:post_levy_ngg} and the
density of the jumps $J_j^*$ is \eqref{postfj}. Finally, the
conditional distribution of $U$, given $\btheta$, is then \eqref{condu}.

The second conditional distribution $[{\btheta}|{\mathbf X},\tilde M]$
involved in the Gibbs sampler in \eqref{cdist} consists of conditional
independent distributions for each $\theta_i$, whose density is given by
%
\begin{equation}
\label{condy} f_{\theta_i|X_i,\tilde M}(s)\propto k(X_i|s) \tilde M^*_+\{s
\}
\end{equation}
for $i=1,\ldots,n$, where the set $\{s\}$ corresponds to the
$m$-variate jump locations $s \in\R^m$ of the posterior process
$\tilde M^*_+$.

In the following we will provide a way of simulating from each of the
distributions \eqref{conda}, \eqref{condu} and~\eqref{condy}.

\subsection{\texorpdfstring{Simulating $[\tilde M|U,\theta]$}{Simulating [M|U, theta]}}

Since the distribution of the process $\tilde M$, given $U$ and~${\btheta}$, is the distribution function associated to a CRM, we need
to sample 
its trajectories. Algorithms for simulating such processes usually rely
on inverse L\'evy measure techniques as is the case for the algorithms
devised in \citet{ferg72} and in \citet{wolpert}. According to \citet
{walker00}, the former is more efficient in the sense that it has a
better performance with a small number of simulations. Therefore, for
simulating from the conditional distribution of $\tilde M$ we follow
the Ferguson and Klass device. Their idea is based on expressing the
part without fixed points of discontinuity of the posterior $\tilde
M_+^*$, which in our case is $\tilde M^*$, as an infinite sum of random
jumps $J_j$ that occur at random locations $\vartheta_j=(\vartheta
_{j}^{(1)},\ldots,\vartheta_{j}^{(m)})$, that is,
%
\begin{equation}
\tilde M^*(s)=\sum_{j=1}^\infty
J_j \indic_{(-\infty, s]}(\vartheta_j).
\label{eq:post_crm_contin}
\end{equation}
The positive random jumps are ordered, that is, $J_1\geq J_2\geq\cdots
$, since the $J_j$'s are obtained as\vadjust{\goodbreak} $\xi_j=N(J_j)$, where $N(v)=\nu
^*([v,\infty),\R^m)$ and $\xi_1,\xi_2,\ldots$ are jump times of a
standard Poisson process of unit rate, that is, $\xi_1,\xi_2-\xi
_1,\ldots\simiid\ga(1,1)$. Here $\ga(a,b)$ denotes a gamma
distribution with shape and scale parameters $a$ and $b$. The random
locations $\vartheta_j$, conditional on the jump sizes $J_j$, are
obtained from the distribution function $F_{\vartheta_j|J_j}$, given by
\[
F_{\vartheta_j|J_j}(s)=\frac{\nu^*(\ddr J_j , (-\infty,s])}{\nu^*(
\ddr J_j,\R^m)}.
\]
Therefore, the $J_j$'s can be obtained by solving the equations $\xi
_i=N(J_i)$. This can be accomplished by combining quadrature methods to
approximate the integral (see, e.g., \cite*{burden}) and a
numerical procedure to solve the equation. Moreover, when one is
dealing with a homogeneous NRMI the jumps are independent of the
locations and, therefore, $F_{\vartheta_j|J_j}=F_{\vartheta_j}$ does
not depend on $J_j$, implying
that the locations are i.i.d. samples from $P_0$. For an extension of the
Ferguson--Klass device to general space see \citet{orbanza}.

In our specific case where $\tilde M$ is a generalized gamma
process, the functions $N$ and $F_\vartheta$ take on the form
%
\begin{eqnarray}
\label{funcM} \qquad N(v)&=&\frac{a}{\Gamma(1-\gamma)}\int_v^\infty
\mathrm{e}^{-(\kappa
+u)x}x^{-(1+\gamma)}\,\d x,
\nonumber
\\[-8pt]
\\[-8pt]
\nonumber
F_\vartheta(s) &=&\int_{(-\infty,s]}
P_0(\d y),
\end{eqnarray}
and all above described steps become straightforward.

As for the part of $\tilde M^*_+$ concerning the fixed points of
discontinuity, the distribution of the jumps at the fixed locations
will depend explicitly on the underlying L\'evy intensity as can be
seen from \eqref{eq:post_jump}. In the NGG case they reduce to the
gamma distributions displayed in \eqref{postfj}.

Now, combining the two parts of the process, with and without fixed
points of discontinuity, the overall posterior representation of the
process $\tilde M$ will be
\[
\tilde M^*_+(s)=\sum_j \bar J_j
\indic_{(-\infty,s]}(\bar\vartheta_j),
\]
having set $\{\bar J_j\}_{j \ge1}=\{J_1^*, \ldots, J_r^*, J_1, \ldots
\}
$ and also $\{\bar\vartheta_j\}_{j \ge1}=\{\theta_1^*, \ldots,
\theta
^*_r, \vartheta_1, \ldots\}$.

\begin{figure*}

\includegraphics{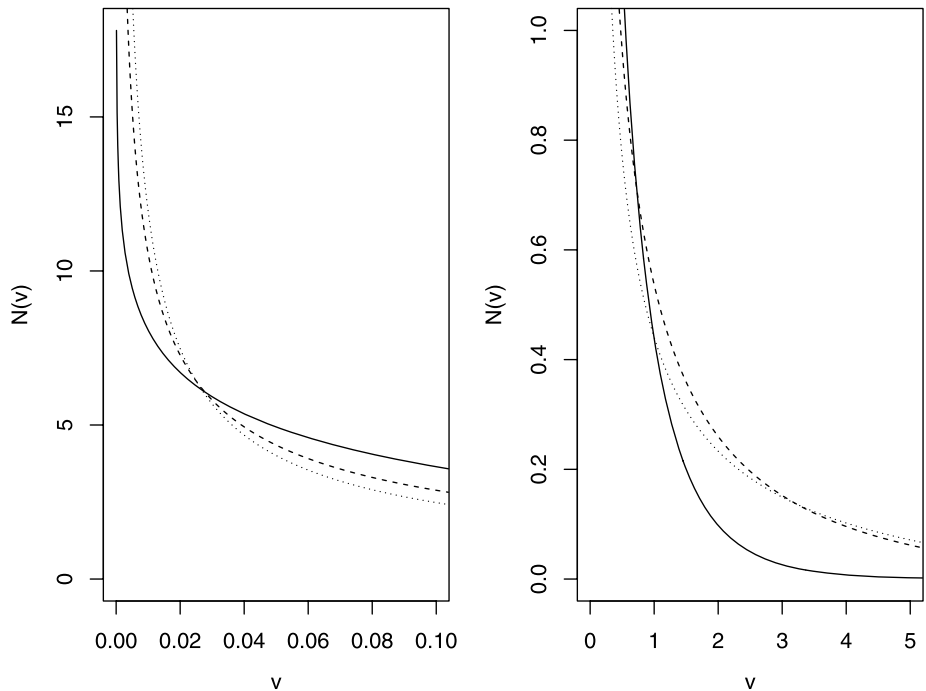}

\caption{Function $N$ in \protect\eqref{funcM} for three
special cases of generalized
gamma processes: gamma process with $(a,\kappa,\gamma)=(2,1,0)$ (solid
line); inverse Gaussian
process with $(a,\kappa,\gamma)= (1,0.126,0.5)$ (dashed line); and
stable process with
$(a,\kappa,\gamma)=(1,0,0.666)$ (dotted line). In all the cases, the
prior mean and variance
of $\tilde{P}(A)$ obtained by normalization are the same, for any $A$.}
\label{figM}
\end{figure*}

\begin{remark} A fundamental merit of Ferguson and Klass'
representation, compared to similar algorithms, is the fact that the
random heights $J_i$ are obtained in a descending order. Therefore, one
can truncate the series \eqref{eq:post_crm_contin} at a certain finite
index $\ell$ in such a way that the relative error between $\sum_{j\leq
\ell}J_j$ and $\sum_{j\leq\ell+1}J_j$ is smaller than $\varepsilon$, for\vadjust{\goodbreak}
any desired \mbox{$\varepsilon>0$}. This, on the one hand, guarantees that the
highest jumps are not left out and, on the other hand, allows us to
control the size of the ignored jumps. \citet{arg2} provide an upper
bound for the ignored jump sizes.
\end{remark}

As mentioned before, the generalized gamma\break NRMI defines a wide class of
processes which include gamma, inverse Gaussian and stable processes.
To appreciate better the difference between these processes, consider
the function $N$ in \eqref{funcM}. This function is depicted in Figure
\ref{figM} for the three cases with parameters fixed in such a way that
the corresponding NRMIs (Dirichlet, normalized inverse Gaussian and
normalized stable) share the same baseline probability measure $P_0 = \alpha/a$ and
have the same mean and variance structures. See
\citet
{lijoi05} and \citet{james1} for the relevant explicit expressions
needed to fix the parameters. In particular, Figure~\ref{figM} is
displayed in two panels which represent close-up views to the upper
left and bottom right tails of the graph.

The function $N$ defines the height of the jumps in the part of the
process without fixed points of discontinuity, that is, $J_i=N^{-1}(\xi
_i)$. To help intuition, imagine horizontal lines going up in Figure~\ref{figM}.
The values in the y-axis\vadjust{\goodbreak} correspond to the Poisson process
jumps and, for each of them, there is a value in the $x$-axis
corresponding to the jump sizes of the process. Looking at the right
panel in Figure~\ref{figM}, we can see that the stable process has the
largest jumps followed closely by the inverse Gaussian process. On the
other hand, the left panel shows the concentration of the sizes of the
jumps of the (unnormalized) CRMs around the origin. Hence, the stable
CRM tends to have a larger number of jumps of ``small'' size when
compared to the Dirichlet process, with the N-IG process again in an
intermediate position. As shown in \citet{kingman}, this different
behavior also impacts the normalized weights. To grasp the idea, let
the $J_i$'s be the jump sizes of the CRM and $\tilde p_i=J_i/\sum_{k\ge
1} J_k$ are the normalized jumps. Moreover, $(\tilde p_{(j)})_{j\ge1}$
is the sequence obtained by considering the $\tilde p_j$'s in
decreasing order so that $\tilde p_{(1)}> \tilde p_{(2)}> \cdots.$ One
then has $\tilde p_{(j)} \sim\exp\{-j/a\} $ as $j\to\infty$, almost
surely, in the Dirichlet case, whereas $\tilde p_{(j)} \sim\xi(\gamma
)  j^{-1/\gamma}$ as $j\to\infty$, almost surely, in the N-stable
case. Here $\xi(\gamma)$ is a positive random variable. Hence, for $j$
large enough the atom $Z_{(j)}$ associated to the weight $\tilde
p_{(j)}$ is less likely to be observed in the Dirichlet case rather
than in the N-stable case. These arguments can be suitably adapted and
the conclusion can be extended to the case where the N-stable is
replaced by a $\operatorname{NGG}(a,\kappa,\gamma,P_0)$ process, for any $\gamma\in
(0,1)$. An important well-known implication of this different behavior
concerns the distribution of the number of distinct values $R_n$:
clearly, for both the Dirichlet and the $\operatorname{NGG}(a,\kappa,\gamma,P_0)$
(with $\gamma>0$) processes $R_n$ diverges as $n$ diverges; however,
the rate at which the number of clusters $R_n$ increases is slower in
the Dirichet than in the NGG case, being, respectively, $\log(n)$ and
$n^\gamma$. Moreover, in order to gain a full understanding of the role
of $\gamma$ in determining the clustering structure featured by models
defined either as in \eqref{eq:nonpmic} or \eqref{eq:mixt}, one has to
consider the influence $\gamma$ has on the sizes of the clusters. To
this end, it is useful to recall that when $\gamma>0$ a reinforcement
mechanism of larger clusters takes place. A concise description is as
follows: Consider a configuration reached after sampling $n$ values,
and denote by $n_i$ and $n_j$ the sizes of the $i$th and $j$th cluster,
respectively, with $n_i>n_j$. Then, the ratio of the probabilities that
the $(n+1)$th sampled value will belong to the $i$th or $j$th clusters
coincides with $(n_i-\gamma)/(n_j-\gamma)$, an increasing function of
$\gamma$, with its lowest value corresponding to the Dirichlet process,
that is, $\gamma=0$. For instance, if $n_i=2$ and $n_j=1$, the
probability of sampling a value belonging to the $i$th cluster is twice
the probability of getting a value belonging to the $j$th cluster in
the Dirichlet case, whereas it is three times larger for $\gamma=1/2$
and five times larger for $\gamma=3/4$. This implies that as $\gamma$
increases, the clusters tend to be much more concentrated with a very
large number of small clusters and very few groups having large
frequencies. In other words, a~mass reallocation occurs and it
penalizes clusters with smaller sizes while reinforcing larger
clusters, which are interpreted as those having stronger empirical
evidence. On the other hand, $\kappa$ (or $a$) does not have any
significant impact on the balancedness of the partition sets. This
mechanism is far from being a drawback and \citet{lijoi07} have shown
that it is beneficial when drawing inference on the number of
components in a mixture. Finally, it is worth stressing that, in
general, the unevenness of partition configurations is an unavoidable
aspect of nonparametric models beyond the specific cases we are
considering here. This is due to the fact that, with discrete
nonparametric priors, $R_n$ increases indefinitely with~$n$. Hence, for
any $n$ there will always be a positive probability that a new value is
generated and, even if at different rates, new values will be
continuously added, making it impossible to obtain models with (a
priori) balanced partitions. If one needs balancedness even a priori, a~finite-dimensional model is more appropriate.

\subsection{\texorpdfstring{Simulating $[U|\theta]$}{Simulating [U|theta]}}

Since the conditional density of $U$ given in \eqref{condu} is
univariate and continuous, there are several ways of drawing samples
from it. \citet{damien}, for instance, propose to introduce uniform
latent variables to simplify the simulation. However, in our
experience, this procedure increases the autocorrelation in the chain,
thus 
leading to a slower \mbox{mixing}. Additionally, the values of this
conditional density explode for sample sizes larger than $100$. An
alternative procedure consists of introducing a Metropolis--Hastings
(M--H) step (see, e.g., \cite*{tierney}). M--H steps usually work
fine as long as the proposal distribution is adequately chosen, and
since they rely only on ratios of the desired density, this solves the
overflow problem for large values of $n$.

In our approach we propose to use a M--H step with proposal
distribution that follows a random walk. Since $U$ takes only positive
values, we use a gamma proposal distribution centered at the previous
value of the chain and with coefficient of variation equal $1/\sqrt {\delta}$. Specifically, at iteration $[t+1]$ simulate $u^\backprime
\sim
\ga(\delta,\delta/u^{[t]})$ and set $u^{[t+1]}=u^\backprime$ with
acceptance probability given by
%
\begin{eqnarray}
\label{apu}&& q_1 \bigl(u^\backprime,u^{[t]} \bigr)
\nonumber
\\[-8pt]
\\[-8pt]
\nonumber
&&\quad=\min
\biggl\{1,\frac{f_{U|{\btheta
}}(u^\backprime)\ga(u^{[t]}|\delta,\delta/u^\backprime)} {
f_{U|{\btheta}}(u^{[t]})\ga(u^\backprime|\delta,\delta
/u^{[t]})} \biggr\},
\end{eqnarray}
where $\ga( \cdot |a,b)$ denotes the density function of a gamma
random variable whose expected value is $a/b$.
The parameter $\delta$ controls the acceptance rate of the M--H step
being higher for larger values.
It is suggested to use $\delta\geq1$.

\subsection{\texorpdfstring{Resampling the Unique Values $\theta_j^*$}{Resampling the Unique Values theta j*}}\label{sec:resampling}
It is well known that discrete nonparametric priors, as is the case of
NRMIs, induce
some effect when carrying out posterior inference via simulation. This
is called by some authors the ``sticky clusters effect.'' \citet{bush}
suggested an important acceleration step to overcome this problem by
resampling the location of the fixed jumps $\theta_j^*$ from its
conditional distribution given the cluster configuration (c.c.), which
in this case takes on the form
%
\begin{equation}
\label{astep} f_{\theta_j^*|{\mathbf X},\mathrm{c.c.}}(s)\propto f_0(s) \prod
_{i\in C_j} k(X_i|s),
\end{equation}
where $C_j=\{i\dvtx \theta_i=\theta_j^*\}$. Also recall that $\theta
_j^*=(\theta_{j1}^{*}, \ldots,\break \theta_{jm}^{*})\in\R^{m}$ with
$m\geq
1$. For the case $m=2$ of location-scale mixture, that is, $\theta
=(\mu
,\sigma)$, we suggest to use a M--H step with joint proposal
distribution for the pair $(\mu,\sigma)$ whose density we denote in
general by~$g$. In particular, at iteration $[t+1]$ one could sample
$\theta^{*\backprime}=(\mu_j^{*\backprime},\sigma_j^{*\backprime
})$ by
first taking $\sigma_j^{*\backprime}\sim\ga(\delta,\delta/\sigma
_j^{*[t]})$ and then, conditionally on $\sigma_j^{*\backprime}$, take
$\mu_{j}^{*\backprime}$ from the marginal base measure on $\mu$,
$f_0^1$, specified in such a way that its mean coincides with $\bar
{X}_j$ and its standard deviation with $\eta\sigma_{j}^{*\backprime
}/\sqrt{n_j}$, where $\bar{X}_j=\break\frac{1}{n_j}\sum_{i\in C_j}X_i$.
Finally, set $\theta_{j}^{*[t+1]}=\theta_{j}^{*\backprime}$ with
acceptance probability given by
%
\begin{eqnarray}
\label{apys} &&q_2 \bigl(\theta^{*\backprime},\theta^{*[t]}
\bigr)
\nonumber
\\[-8pt]
\\[-8pt]
\nonumber
&&\quad= \min \biggl\{1, \frac{f_{\theta_j^*|{\mathbf X},\mathrm{c.c.}}(\theta^\backprime)} {
f_{\theta_j^*|{\mathbf X},\mathrm{c.c.}}(\theta^{*[t]})} \frac{g(\theta^{*[t]})} {
g(\theta^{\backprime})} \biggr\}.
\end{eqnarray}
For the examples considered in this paper we use $\delta=4$ and $\eta
=2$ to produce a moderate acceptance probability.

\subsection{\texorpdfstring{Simulating $[{\theta}|{X},\tilde M]$}{Simulating [theta|X, M]}}

Since $\tilde M^*_+$ is a pure jump process, the support of the
conditional distribution of $\theta_i$ are the locations of the jumps
of $\tilde M^*_+$, that is, $\{\bar\vartheta_j\}$, and, therefore,
%
\begin{equation}
\label{cdensy} f_{\theta_i|X_i,\tilde M}(s)\propto\sum_j
k(X_i|s)\bar J_j\delta _{\bar
\vartheta_j}(\ddr s).
\end{equation}
Simulating from this conditional distribution is\break straightforward: one
just needs to evaluate the right-hand side of the expression above and
normalize.

\subsection{Updating the Hyperparameters of $P_0$} \label{sec:phi}
As pointed out by one of the referees, in general the hyperparameters
$\phi$ of the base measure density $f_0(  \cdot  |\phi)$ affect the
performance of nonparametric mixtures. For the location-scale mixture
case, that is, $m=2$ with $\theta=(\mu,\sigma)$ and $f_0(\mu,\sigma
|\phi
)=f_{0}^1(\mu|\sigma,\break\varphi)f_{0}^2(\sigma|\varsigma)$, it turns out
that the subset of parameters $\varphi$ pertaining to the locations
$\mu
$ have a higher impact. By assuming in addition a priori independence
between $\mu$ and $\sigma$, the conditional posterior distribution of
$\varphi$, given the observed data and the rest of the parameters, only
depends on the distinct $\mu_i$'s, say, $\mu_j^*$, for $j=1,\ldots,r$.
The simplest way to proceed is to consider a conjugate prior $f(\varphi
)$ for a sample $\mu_1^*,\ldots,\mu_r^*$ from $f_0^1(\mu|\varphi)$.
Clearly such a prior depends on the particular choice of $f_0^1$ and
some examples will be considered in Section~\ref{sec:illustration}.

\subsection{\texorpdfstring{Computing a Path of $\tilde{f}(x)$}{Computing a Path of f(x)}}

Once we have a sample from the posterior distribution of the process
$\tilde M$, the desired path from the posterior distribution of the
random density $\tilde{f}$, given in
\eqref{eq:mixt}, can be expressed as a discrete mixture of the form
%
\begin{equation}
\label{pathf} \tilde f \bigl(x|\tilde M^*_+,\phi \bigr)=\sum
_j k(x|\bar\vartheta_j)\frac
{\bar
J_j}{\sum_l \bar J_l}.
\end{equation}

\subsection{General Algorithm}\label{sec3.7}
An algorithm for simulating from the posterior~distributions \eqref
{cdist} can be summarized as follows. Given the starting points $\theta
_1^{[0]},\ldots,\theta_n^{[0]}$, with the corresponding unique values
$\theta_j^{*[0]}$ and frequencies $n_j^{[0]}$,
for $j=1,\ldots,r$, and given $u^{[0]}$, 
at iteration $[t+1]$:
\begin{enumerate}
\item
Sample the latent $U|{\btheta}$: simulate a proposal value
$U^*\sim\ga(\delta,\delta/U^{[t]})$ 
and take $U^{[t+1]}=U^*$ with probability $q_1(U^*,U^{[t]})$, otherwise take
$U^{[t+1]}=U^{[t]}$, where the acceptance probability $q_1$ is given in~\eqref{apu}.
\item
Sample trajectories of the part of the process without fixed points of
discontinuity
$\tilde M^*$: simulate $\zeta_j\sim\ga(1,1)$ and find $J_j^{[t+1]}$
by solving
numerically the equation $\sum_{l=1}^j\zeta_l=N(J_j)$; simulate
$\vartheta_j^{[t+1]}$
from $P_0$. The function $N$ is given in~\eqref{funcM}. Stop simulating
when $J_{\ell+1}/\sum_{j=1}^\ell J_j<\varepsilon$, say, $\varepsilon=0.0001$.
\item
Resample the unique values $\{\theta_j^*\}$: record the unique values
$\theta_j^{*[t]}$
from $\{\theta_1^{[t]},\ldots,\theta_n^{[t]}\}$ and their frequency
$n_j^{[t]}$.\vspace*{1pt} If $m=2$ with $k(  \cdot |\theta)$ parameterized in
terms of mean and standard deviation ($\theta=(\mu,\sigma)$), simulate
a pair $(\mu_j^{*\backprime},\sigma_j^{*\backprime})$ from a joint
proposal (see Section~\ref{sec:resampling})
and then set
$\theta_{j}^{*[t+1]}$ equal to $\theta_{j}^{*\backprime}$ with
probability $q_2(\theta^{*\backprime},\theta^{*[t]})$. Otherwise take
$\theta_{j}^{*[t+1]}=\theta_{j}^{*[t]}$. The acceptance probability
$q_2$ is given in~\eqref{apys}.
\item
Sample the fixed jumps of the process, $\{J_j^*\}$: for each 
$\theta_j^{*[t+1]}$ with frequency $n_j^{[t+1]}$, $j=1, \ldots, r$,
sample the jump $J_j^{*[t+1]}\sim\ga(n_j^{[t+1]}-\gamma,\kappa+u^{[t+1]})$.
\item
Update the hyperparameters $\bphi$ of $f_0(\theta|\bphi)$: in
particular, for the case of $m=2$ with $\theta=(\mu, \sigma)$ simulate
a value $\bolds{\varphi}^{[t+1]}$ from its conditional posterior
distribution as described in Section~\ref{sec:phi}.
\item
Sample the latent vector ${\btheta}$: for each $i=1,\ldots,n$, sample
$\theta_i^{[t+1]}$
from its discrete conditional density given in \eqref{cdensy} by
evaluating the
kernel $k(X_i| \cdot )$ at the different jump locations\vadjust{\goodbreak} $\{\bar
\vartheta_j^{[t+1]}\}=\{\theta_1^{*[t+1]},\break\ldots, \theta
_r^{*[t+1]},\vartheta_j^{[t+1]},\ldots\}$
and weights $\{\bar J_j^{[t+1]}\}=\break\{J_1^{*[t+1]}, \ldots,  J_r^{*[t+1]},
J_1^{[t+1]}, \ldots\}$.
\item
Compute a path of\vspace*{1pt} the desired random density function $\tilde
f(x|(\tilde M^*_+)^{[t+1]})$ as in \eqref{pathf}.
\end{enumerate}
Repeat steps 1 to 7 for $t=1,\ldots,T$. Note that the values of
$\delta
$ and $\eta$
can be used to tune the acceptance probability in the M--H steps. The
values suggested
here are those considered more appropriate according to our experience.
The performance of this algorithm depends on the particular choices of
the density kernel, the NRMI driving measure and the data set at hand.
In order to assess the mixing of the chains, one can resort to the
effective sample size (ESS) implemented in the {\textsf R} package
library \texttt{coda}. In our context the natural parameter to consider
for assessing the mixing is given by the total jump sizes of the NRMI
process $\sum_{j}\bar{J}_j$. First note that the conjugacy of the
Dirichlet process yields a simpler posterior representation
(independent of the latent variable $U$) and recall also that the jumps
are independent of the locations. Therefore, the samples are
independent and the ESS coincides with the number of iterations of the
chain. For the other NRMIs this is not the case: the posterior
representation depends on the latent variable $U$ and, moreover, the
distribution of the jumps depends on the $\theta_j^*$'s. For instance,
for the two real data sets considered in Section~\ref{sec4.1}, for chains of
length 4500 (obtained from 20,000 iterations with burn-in of 2000
and keeping every 4th iteration), the ESS was around 1250 for the N-IG
process and for the associated latent variable $U$, the value of the
ESS was 1500.

\section{Comparing NRMI Mixtures}
\label{sec:illustration}

In this section we provide a comprehensive illustration of NRMI
mixtures using the {\textsf R} package\break \texttt{BNPdensity}, which
implements the general algorithm outlined in Section~\ref{sec3.7}. The aim of
such a study is twofold: on the one hand, it illustrates the potential
and flexibility of NRMI mixture models in terms of fitting and
capturing the appropriate number of clusters in a data set for
different choices of kernels and mixing NRMI; on the other hand, we
also compare the performance of NRMI mixtures with respect to other
alternative density estimates.

To implement the algorithm described in the previous section, we first
specify the mixture kernel $k(\cdot|\theta)$. We will consider, in
total, a set of four kernels parameterized in terms of mean $\mu$ and
standard deviation $\sigma$ such that $\theta=(\mu,\sigma)$. Two of
these kernels have support $\R$ and the other two have support $\R^+$.
They are as follows:
\begin{longlist}[(iii)]
\item[(i)] Normal kernel:
\[
k(x|\mu,\sigma)=\frac{1}{\sqrt{2\pi}b}\exp \biggl\{-\frac{1}{2
b^2}(x-a)^2
\biggr\}\indic_{\R}(x),
\]
with $a=\mu$ and $b=\sigma$.
\item[(ii)] Double exponential kernel:
\[
k(x|\mu,\sigma)=\frac{1}{2b}\exp \biggl\{-\frac{1}{b}|x-a| \biggr\}
\indic _{\R}(x),
\]
with $a=\mu$ and $b=\sigma/\sqrt{2}$.
\item[(iii)] Gamma kernel:
\[
k(x|\mu,\sigma)=\frac{b^a}{\Gamma(a)}x^{a-1}\mathrm{e}^{-bx}
\indic_{\R^+}(x),
\]
with $a=\mu^2/\sigma^2$ and $b=\mu/\sigma^2$.
\item[(iv)] Log-normal kernel:
\[
k(x|\mu,\sigma)=\frac{1}{x\sqrt{2\pi}b}\exp \biggl\{-
\frac
{1}{2 b^2}(\log x-a)^2 \biggr\}\indic_{\R^+}(x),
\]
with $a=\log (\frac{\mu}{\sqrt{1+\sigma^2/\mu^2}} )$ and
$b=\sqrt{\log (1+\frac{\sigma^2}{\mu^2} )}.$
\end{longlist}

As for the NRMI mixing measure, we will resort to different members of
the class $\operatorname{NGG}(a, \kappa, \gamma;$ $P_0)$: the
Dirichlet process
$\operatorname{NGG}(a, 1, 0; P_0)$, the N-IG process $\operatorname
{NGG}(1, \kappa,
1/2; P_0)$, the N-stable process $\operatorname{NGG}(1, 0,\break \gamma;$ $P_0)$.
Their parameters will be fixed to obtain mixtures with a prior expected
number of components $\E(R_n)$ equal to any desired number $c \in\{1,
\ldots, n\}$, where $n$ denotes the sample size. This strategy allows
one to effectively compare different priors given they induce a priori
the same expected number of mixture components. See \citet{lijoi07} for
details on this procedure. As for the base measure $P_0$ of the NRMIs
to be considered, we will assume a priori independence between $\mu$
and $\sigma$ so that $f_0(\mu,\sigma|\phi)=f_{0}^1(\mu|\varphi
)f_{0}^2(\sigma| \varsigma)$. In particular, we will take
$f_0^2(\sigma
|\varsigma)=\ga(\sigma|\varsigma_1,\varsigma_2)$, with shape
$\varsigma
_1$ and scale $\varsigma_2$ fixed a priori to specify a certain
knowledge in the degree of smoothness. For $f_0^1$ we will consider two
options with support $\R$ and $\R^+$, respectively. These are as follows:
\begin{longlist}[(a)]
\item[(a)] Normal base measure for $\mu$:
\[
f_0^1(\mu|\varphi)=\no(\mu|\varphi_1,
\varphi_2),
\]
where $\varphi_1$ and $\varphi_2$ are the mean and precision, respectively.
The conjugate prior distribution for $\varphi$ is then $f(\varphi
)=\no
(\varphi_1|\psi_1,\psi_2 \varphi_2)\ga(\varphi_2|\psi_3,\psi
_4)$ and\vadjust{\goodbreak}
the (conditional) posterior distribution, needed for the hyperparameter
updating (see Section~\ref{sec:phi}), are given by
\begin{eqnarray*}
f \bigl(\varphi|\mu^* \bigr)&=&\no \biggl(\varphi_1
\Big\vert \frac
{\psi_2\psi
_1+r\bar{\mu}^*}{\psi_2+r}, (\psi_2+r)\varphi_2
\biggr) \\
&&{}\cdot
\ga \Biggl(\varphi_2\Big\vert \psi_3+
\frac{r}{2}, \psi _4+\frac{1}{2}\sum
_{j=1}^{r} \bigl(\mu_j^*-\bar{\mu}^*
\bigr)^2\\
&&\hspace*{95pt}{}+\frac{\psi_2
r(\bar
{\mu}^*-\psi_1)^2}{2(\psi_2+r)} \Biggr).
\end{eqnarray*}
\item[(b)] Gamma base measure for $\mu$:
\[
f_0^1(\mu|\varphi)=\ga(\mu|1,\varphi),
\]
where $\varphi$ corresponds to the scale parameter. The conjugate prior
for $\varphi$ is $f(\varphi)=\ga(\varphi|\psi_1,\psi_2)$ and the
(conditional) posterior distribution is $f(\varphi|\mu^*)=\ga
(\varphi
|\psi_1+r, \psi_2+\sum_{j=1}^r\mu_j^*)$. Clearly, this choice is
reasonable only for experiments leading to positive outcomes.
\end{longlist}

Since we aim at comparing the performance of NRMI mixtures in terms of
density estimates, we also need to specify measures of goodness of fit.
We will use two different measures for the real data and the simulated
data. In the former case, we resort to the conditional predictive
ordinates (CPOs) statistics, which are now widely used in several
contexts for model assessment. See, for example, \citet{gelfand1}.
For each observation $i$, the CPO statistic 
is defined as follows:
\[
\operatorname{CPO}_i=\tilde{f} \bigl(x_i|D^{(-i)}
\bigr)=\int k(x_i|\theta)\tilde P \bigl(\d \theta |D^{(-i)}
\bigr),
\]
where $D^{(-i)}$ denotes the observed sample $D$ with the $i$th case excluded and $\tilde
P(\d\theta|D^{(-i)})$ the posterior density of the model parameters
$\theta$ based on data $D^{(-i)}$. By rewriting the statistic $\operatorname
{CPO}_i$ as
\[
\operatorname{CPO}_i= \biggl(\int\frac{1}{k(x_i|\theta)}\tilde P(d\theta |D)
\biggr)^{-1},
\]
it can be easily approximated by Monte Carlo as
\[
\widehat{\operatorname{CPO}_i}= \Biggl(\frac{1}{T}\sum
_{t=1}^T \frac
{1}{k(x_i|\theta^{[t]})} \Biggr)^{-1},
\]
where $\{\theta^{[t]}, t=1,2,\ldots,T\}$ is an MCMC sample from
$\tilde P(\theta|D)$. We will summarize the CPO$_i$, $i=1,\ldots,n$,
values in two ways, as an average of the logarithm of CPOs (ALCPO) and
as the median of the logarithm of CPOs (MLCPO). The average of log-CPOs
is also called the average of log-pseudo marginal likelihood and is
denoted by ALPML.\vadjust{\goodbreak}

In contrast, when considering simulated data, the true model, say,
$f^\ast$, is known and, hence, it is possible to use the mean
integrated squared error (MISE) for model comparison. If we denote by
$\hat{f}_n$ the density estimate conditional on a sample of size $n$
from $f^\ast$, then the MISE
is defined as
\[
\operatorname{MISE}=\E \biggl\{\int \bigl\{\hat{f}_n(x)-f^*(x) \bigr
\}^2\,\d x \biggr\}.
\]
Like in other approaches to density estimation (see, e.g., \cite*{MV}; \cite*{roeder2}),
the standard method to compare with is the kernel
density estimator \citep{silver}. Therefore, instead of the MISE, we
report the relative MISE (RMISE) defined as the ratio of the MISE
obtained with the NRMI mixture model and the MISE obtained with the
kernel density estimator with standard bandwidth.

We are now in a position to illustrate our methodology. We first
provide the analysis of two real data sets popular in the mixture
modeling literature, name\-ly, the galaxy data and the enzyme data. See
\citet{rich}. Then, we perform an extensive simulation study by
considering the models dealt with in \citet{marron}. In analyzing the
real data we focus on the performance of different NRMI mixtures, by
varying kernel and mixing NRMI, and illustrate the flexibility of the
algorithm. Later, through the simulation study we aim at comparing NRMI
mixtures with other methods used in the literature. For this purpose we
fix a single NRMI mixture. Such a choice, based on the results of the
real data examples and on our previous experience, exhibits good and
robust performances, thus making it a valid default model.

\begin{table*}
\caption{Galaxy data set: Summaries of log-conditional
predictive ordinates [average (ALCPO) and median (MLCPO)] and mode of
the posterior distribution of the number of components in the mixture,
$R_n|{\mathbf{X}}$, for different prior specifications. Bold numbers denote
best fitting according to the corresponding statistic}\label{table1}
\begin{tabular*}{\textwidth}{@{\extracolsep{\fill}}lccccc@{}}
\hline
\textbf{Measure} & \textbf{Kernel} & \multicolumn{1}{c}{$\bolds{(\varsigma_1,\varsigma_2)}$} & \textbf{ALCPO} & \textbf{MLCPO} &
\multicolumn{1}{c@{}}{\textbf{Mode}$\bolds{(R_n|{X})}$} \\
\hline
Dirichlet & Normal & $(1,1)$ & $\bolds{-2.581}$ & $-2.250$ & 7 \\
& & $(0.1,0.1)$ & $-2.619$ & $-2.205$ & 6 \\
& Dble.Exp. & $(1,1)$ & $-2.597$ & $-2.303$ & 7 \\
& & $(0.1,0.1)$ & $-2.620$ & $-2.305$ & 6 \\[3pt]
N-IG & Normal & $(1,1)$ & $-2.608$ & $\bolds{-2.099}$ & 5 \\
& & $(0.1,0.1)$ & $-2.647$ & $-2.154$ & 3 \\
& Dble.Exp. & $(1,1)$ & $-2.600$ & $-2.258$ & 5 \\
& & $(0.1,0.1)$ & $-2.637$ & $-2.260$ & 4 \\
\hline
\end{tabular*}
\end{table*}

\subsection{Real Data}\label{sec4.1}

\subsubsection{Galaxy data}
For illustration of the algorithm and analysis of NRMI mixtures we
start with some real data. The first data set we consider is the widely
studied galaxy data set. Data
consist of velocities of $82$ distant galaxies diverging from our own
galaxy. Typically this density has been estimated by considering
mixtures of normal kernels (\cite*{escobar}; \cite*{rich}; \cite*{lijoi05}): given the
data range from $9.2$ to $34$, clearly away from zero, it is possible
to use kernels with support $\R$. Here, we compare the normal kernel
with another kernel with real support, namely, the double exponential
kernel. These two kernels are written in mean and\vadjust{\goodbreak} standard deviation
parameterization as in cases (i) and (ii) above. In terms of mixing
measures we compare two options: the Dirichlet process with
specifications $\operatorname{NGG}(3.641,1,0; P_0)$ and the N-IG
process with
specifications $\operatorname{NGG}(1,0.015,1/2; P_0)$. The prior
parameters of
the two processes were
determined so as to obtain an expected number of a priori components
equal to $12$, roughly twice the typically estimated number of
components, which is between $4$ and $6$. It is worth noting that with
such a prior specification the N-stable process would correspond to a
$\operatorname{NGG}(1,0,0.537; P_0)$. This essentially coincides with
the above
N-IG specification which indeed has a small value of $\kappa$ and
$\gamma=1/2$, and is therefore omitted.

For the base measure $P_0$ we took $f_0^2(\sigma|\varsigma)= \ga
(\sigma
|\varsigma_1,\break\varsigma_2)$ with two specifications for $(\varsigma
_1,\varsigma_2)$, namely, $(1,1)$ and $(0.1,0.1)$, and the gamma
specification in case (b) above for $f_0^1(\mu|\varphi)$ with a vague
hyperprior on the scale parameter $\varphi$, namely, $\psi_1=\psi
_2=0.01$. In neither case $P_0$ is conjugate w.r.t. the kernel and in
addition to the standard deviations, it forces also the means of the
mixture components to be positive as required. The Gibbs sampler was
run for $20\mbox{,}000$ iterations with a burn-in of $2000$ sweeps. One
simulation every $4$th after burn-in was kept, resulting in $4500$
iterations to compute the estimates.

\begin{figure*}

\includegraphics{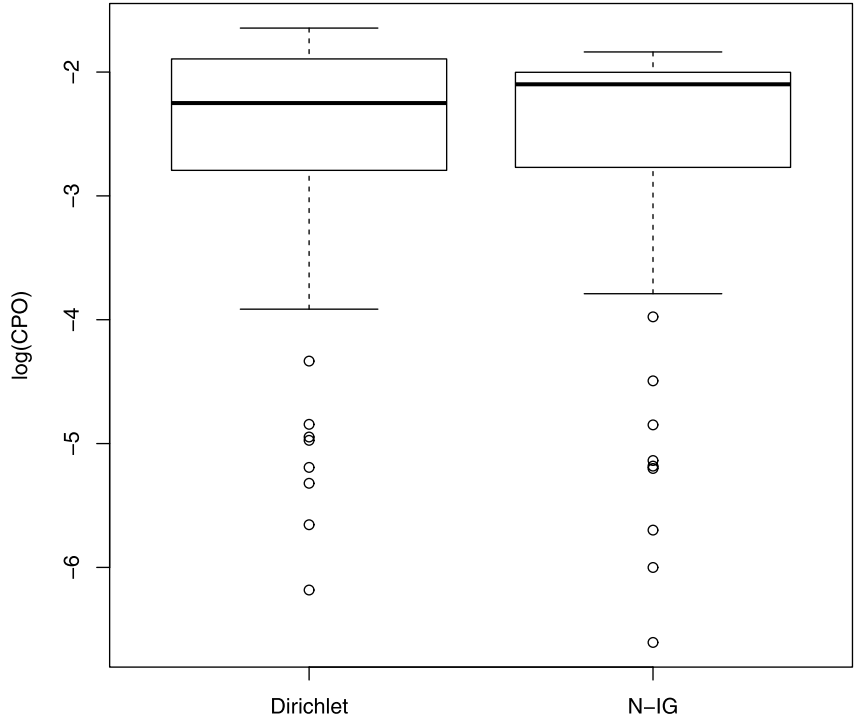}

\caption{Galaxy data set: Box-plot of the logarithm of the conditional
predictive
ordinates for DPM and N-IG mixtures, with normal kernel and
$(\varsigma
_1,\varsigma_2)=(1,1)$.}
\label{cpo1}
\end{figure*}

Table~\ref{table1} provides the ALCPO statistics, the MLCPO statistics
and the mode of posterior distribution of the number of components,
$R_n|{\mathbf X}$, for the $8=2\times2\times2$ combinations of
kernel-NRMI-$(\varsigma_1,\varsigma_2)$. Recall that the ALCPO and
MLCPO statistics are the average and the median of the CPOs in log
scale, respectively. First note that starting from an ``incorrect''
prior specification of the number of components $R_n$, the N-IG
process mixture is able to detect the typically estimated number of
components regardless of the choice of the kernel and the other
parameters. In contrast, DPMs are not able to overcome completely the
wrong prior specification and tend to overestimate the number of
components. As one would expect, given, on the one hand, a distribution
can always be fitted with more components than necessary and, on the
other, the kernel smooths out differences in the mixing measures, the
differences between the two processes in terms of the density estimates
are much less evident. Considering the ALCPO goodness-of-fit
statistics, the best fitting is obtained with the normal DPM with
$(\varsigma_1,\varsigma_2)=(1,1)$. However, the differences
w.r.t. other specifications are not particularly remarkable. If,
instead, we consider the MLCPO statistic, the best fitting
is achieved by the N-IG normal mixture with $(\varsigma_1,\varsigma
_2)=(1,1)$ and the superior performance starts becoming significant,
being $0.1$ better than any DPM specification. The overall behavior of
the CPO is illustrated by Figure~\ref{cpo1}, where box-plots of the
logarithm of the CPO values corresponding to normal mixtures with
$(\varsigma_1,\varsigma_2)=(1,1)$ for both Dirichlet and N-IG processes
are depicted. Coherently with the values of the ALCPO and MLCPO, the
logarithm of the CPOs produced by the DPM are more dispersed: for some
trajectories it produces the best ordinates, which, once averaged, lead
to a slightly better ALCPO; however, if we consider a more robust
summary, like the median, the N-IG mixture produces a significantly
better result.

\begin{figure*}[t!]

\includegraphics{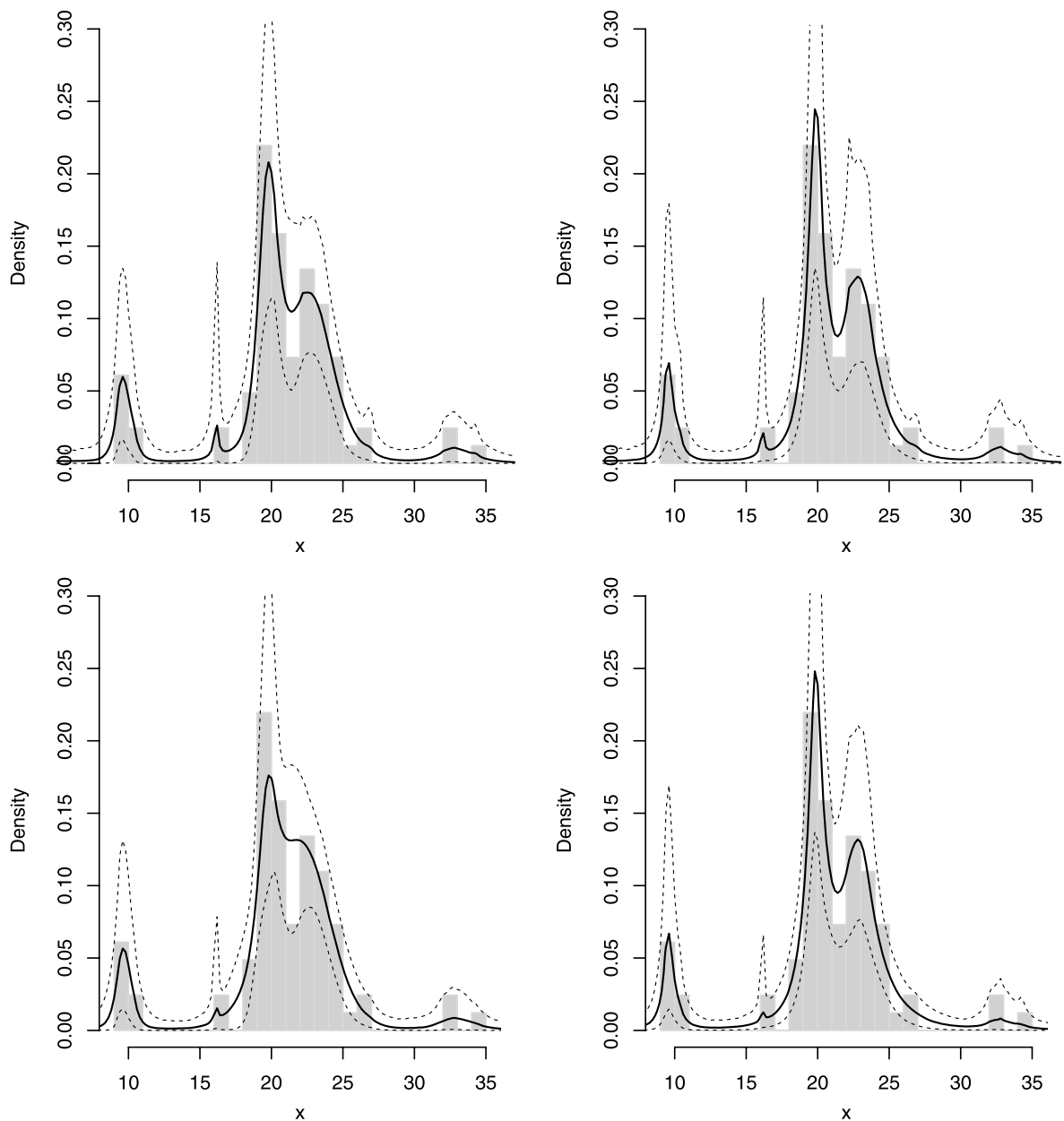}

\caption{Galaxy data set: Posterior density estimates with $(\varsigma
_1,\varsigma_2)=(1,1)$
corresponding to the DPM (top row) and the N-IG mixture (bottom row)
with normal kernel (left column)
and double exponential kernel (right column).}
\label{dex1nig2}\vspace*{-3pt}
\end{figure*}

Figure~\ref{dex1nig2} displays the density estimates together with 95\%
pointwise credible intervals when using the Dirchlet and N-IG process
mixtures with normal and double exponential kernels. In accordance to
the above results, there is not much difference in terms of the chosen
nonparametric prior. However, it is interesting to note how the double
exponential kernel, while exhibiting poorer performance in terms of
CPO, produces significantly sharper estimates than the normal kernel.
This feature which singles out possible modes may be desirable in
certain situations.

\subsubsection{Enzyme data}
The second example consists of $245$ measurements of the enzymatic
activity in the blood of unrelated patients. The values of this data
set are all positive and close to zero, ranging from $0.021$ to $2.9$.
\citet{rich} analyzed this data set and applied a finite mixture of
normals model to estimate the density, even though the data are fairly
close to zero. Instead of working with real support kernels, we perform
our analysis with positive support kernels to be more consistent with
the nature of the data. In particular, we take the gamma density kernel
and the log-normal density kernel, both with the mean and standard
deviation parameterizations as displayed in cases (iii) and (iv) at the
beginning of the section.

As for the nonparametric mixing measures, we consider the Dirichlet
process $\operatorname{NGG}(4.977,1,0;$ $P_0)$ and the N-IG process
$\operatorname
{NGG}(1,0.007,1/2; P_0)$. The prior parameters were fixed so as to
obtain an expected number of a priori components equal to $20$.
Again, the specification of the corresponding N-stable process
$\operatorname
{NGG}(1,0,0.523; P_0)$ essentially coincides with the above N-IG
process and is therefore omitted. Note that such a value for the prior
expected number of components is much larger than the typically $2$ or
$3$ components estimated for this data set. As for the base measure
$P_0$, we took $f_0^2(\sigma|\varsigma)=\ga(\sigma|\varsigma
_1,\varsigma
_2)$ with two possible sets of values for the hyperparameters, that is,
$(\varsigma_1,\varsigma_2)=(4,1)$ and $(\varsigma_1,\varsigma
_2)=(0.5,0.5)$. Moreover, for $\mu$ the gamma specification in (b) is
adopted with a vaguely informative hyperprior on the scale, namely,
$\psi_1=\psi_2=0.01$. We remark that, as in the previous example, these
choices give rise to base measures that are not conjugate for the
kernel. The Gibbs sampler was run for $20\mbox{,}000$ iterations with a
burn-in of $2000$ sweeps, keeping one simulation of every 4th, ending up with $4500$ iterations to compute the
estimates.

\begin{table*}
\caption{Enzyme data set: Summaries of log-conditional
predictive ordinates [average (ALCPO) and median (MLCPO)] and mode of
the posterior distribution of the number of components in the mixture,
$R_n|{\mathbf X}$, for different prior specifications. Bold numbers denote
best fitting according to the corresponding statistic}\label{table2}
\begin{tabular*}{\textwidth}{@{\extracolsep{\fill}}lccccc@{}}
\hline
\textbf{Measure} & \textbf{Kernel} & $\bolds{(\varsigma_1,\varsigma_2)}$ & \textbf{ALCPO} & \textbf{MLCPO} &
\multicolumn{1}{c@{}}{\textbf{Mode}$\bolds{(R_n|\mathbf{X})}$} \\
\hline
Dirichlet & Gamma & $(4,1)$ & $-0.227$ & 0.204 & \phantom{0}5 \\
& & $(0.5,0.5)$ & $-0.218$ & 0.126 & 13 \\
& Log.N. & $(4,1)$ & $-0.216$ & 0.054 & \phantom{0}8 \\
& & $(0.5,0.5)$ & $\bolds{-0.205}$ & 0.006 & 14 \\[3pt]
N-IG & Gamma & $(4,1)$ & $-0.217$ & $\bolds{0.275}$ & \phantom{0}2 \\
& & $(0.5,0.5)$ & $-0.213$ & 0.233 & \phantom{0}5 \\
& Log.N. & $(4,1)$ & $-0.210$ & 0.065 & \phantom{0}5 \\
& & $(0.5,0.5)$ & $-0.208$ & 0.048 & \phantom{0}8 \\ 
\hline
\end{tabular*}
\end{table*}

\begin{figure*}[t!]

\includegraphics{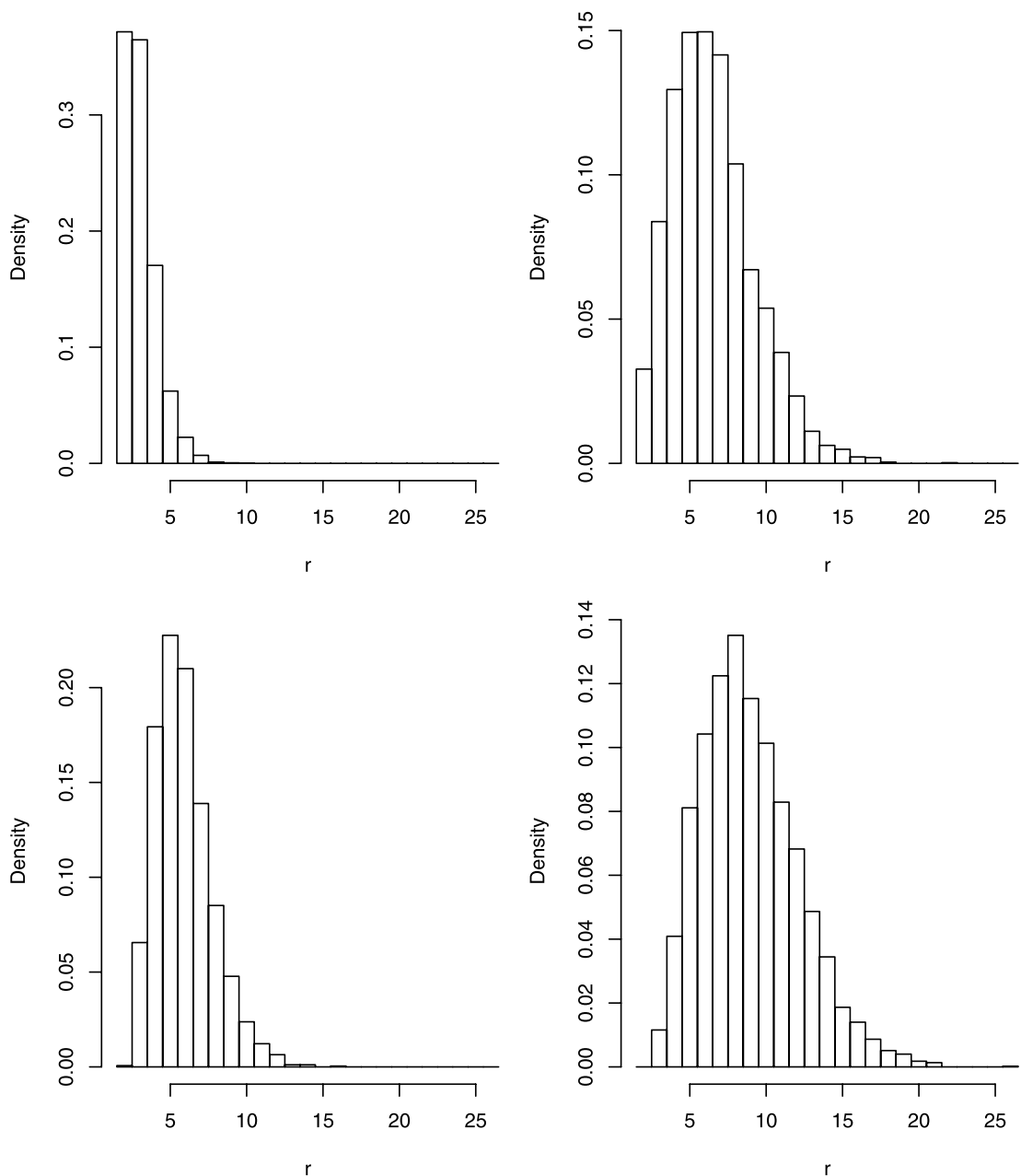}

\caption{Enzyme data set: Posterior distribution for the number of
components, $R_n|{\mathbf X}$, for the N-IG process mixture:
gamma kernel (top row) and log-normal kernel (bottom row) with
$(\varsigma_1,\varsigma_2)=(4,1)$ (left column) and
$(\varsigma_1,\varsigma_2)=(0.5,0,5)$ (right column).}
\label{rex2nig}
\end{figure*}

Table~\ref{table2} provides the ALCPO statistics, the MLCPO statistics
and the mode of the posterior distribution of the number of components
for the $8=2\times2\times2$ combinations of kernel-NRMI-$(\varsigma
_1,\varsigma_2)$, respectively. Let us first focus on the estimated
number of components. In this case, starting from a ``strongly
incorrect'' prior specification of the number of components, the
ability of N-IG mixtures to overcome misspecifications becomes even
more apparent. Indeed, it can be seen that the N-IG mixture estimates
at least 3 fewer components than the DPM, for any choice of the kernels
and of the base measures hyperparameters. Having established the better
performance of the N-IG mixtures, we have a closer look at the impact
of the kernels and hyperparameter specifications in Figure \ref
{rex2nig}. We display the corresponding complete posterior
distributions of the number of components. The gamma kernel displays a
better performance in locating the number of components with,
additionally, a lower variability, regardless of the hyperparameters
choice. With respect to the choice of hyperparameters in the
distribution of $\sigma$, the ones generating larger values with higher
variability are superior. When looking at the density estimates the
differences are, as in the previous example, less apparent. In terms of
the ALCPO goodness-of-fit statistics, the best fitting is obtained
through the DPM with lognormal kernel and $(\varsigma_1,\varsigma
_2)=(0.5,0.5)$, but the differences with respect to the other
specifications are minimal. Nonetheless, it is worth pointing out that
this corresponds to the case which has the worst behavior in terms of
estimation of the number of components. On the one side, this confirms
that using more components than necessary does not impact the fit in
terms of density estimation. On the other hand, it represents an
indication that goodness-of-fit summaries have to be handled with some
care to understand the numerical output. If we consider the MLCPO
statistic, the best fitting is achieved by the model one would actually
expect on the basis of the analysis of the posterior distribution of
the number of components, namely, the N-IG process mixture with gamma
kernel and $(\varsigma_1,\varsigma_2)=(4,1)$. Moreover, its superiority
is quite significant w.r.t. all other specifications. This enforces our
previous comment concerning the care needed in drawing conclusions from
numerical summaries of the fit.

\subsection{Simulation Study}
We now provide an extensive simulation study and use it also for
comparing the performance of NRMI mixtures with other density
estimation methods. \citet{marron} considered a set of $15$ densities
with different behaviors, which are challenging to estimate. These
densities are either unimodal, multimodal, symmetric and/or skewed.
According to \citet{marron}, the last 5 densities are strongly
multimodal and are difficult to recover with moderate sample sizes.
Therefore, we concentrate on their first 10 densities to test the
performance of NRMI mixtures. For each of the $10$ models, the
simulation study was based on $N=40$ simulation experiments and for
each experiment a sample of size $n=250$ was drawn from the
model.\vadjust{\goodbreak}

We considered NRMI mixtures with a normal kernel (i) and a N-stable
process $\operatorname{NGG}(1,0,0.396; P_0)$ as mixing measure. This
choice of
the parameter $\gamma=0.396$ implies that the a priori expected number
of components is equal to $10$, which seems a reasonable default choice.
As for the base measure $P_0$, we took $f_0^2(\sigma|\varsigma)=\ga
(\sigma|1,1)$, whereas for $\mu$ we adopted the normal specification in
(a). As for the latter, the hyperparameters of the normal-gamma prior
on $(\varphi_1, \varphi_2)$ are $\psi_1=0$, $\psi_2=0.01$, $\psi_3=0.1$
and $\psi_4=0.1$.
It is important to note that these prior specifications were the same
for all $10$ models and, hence, all experiments: the idea is to verify
its performance as a default choice rather than tailoring the model on
each specific example. As we mentioned at the beginning of the section,
since these are simulation experiments, one can compute the relative
mean integrated squared error (RMISE) as a measure of goodness of fit.
As benchmarking nonparametric kernel density estimator, w.r.t. which
the RMISE is computed, we considered the optimal bandwidth given in
\citet{silver} which is $\sigma=s^2(1.06)^2n^{-2/5}$, with $s^2$ being
the sample variance. For each case the Gibbs sampler was run for
$10\mbox{,}000$ iterations with a burn-in of $1000$ sweeps and one simulation
every 4th was taken for computing the estimates.

\begin{figure*}[t!]

\includegraphics{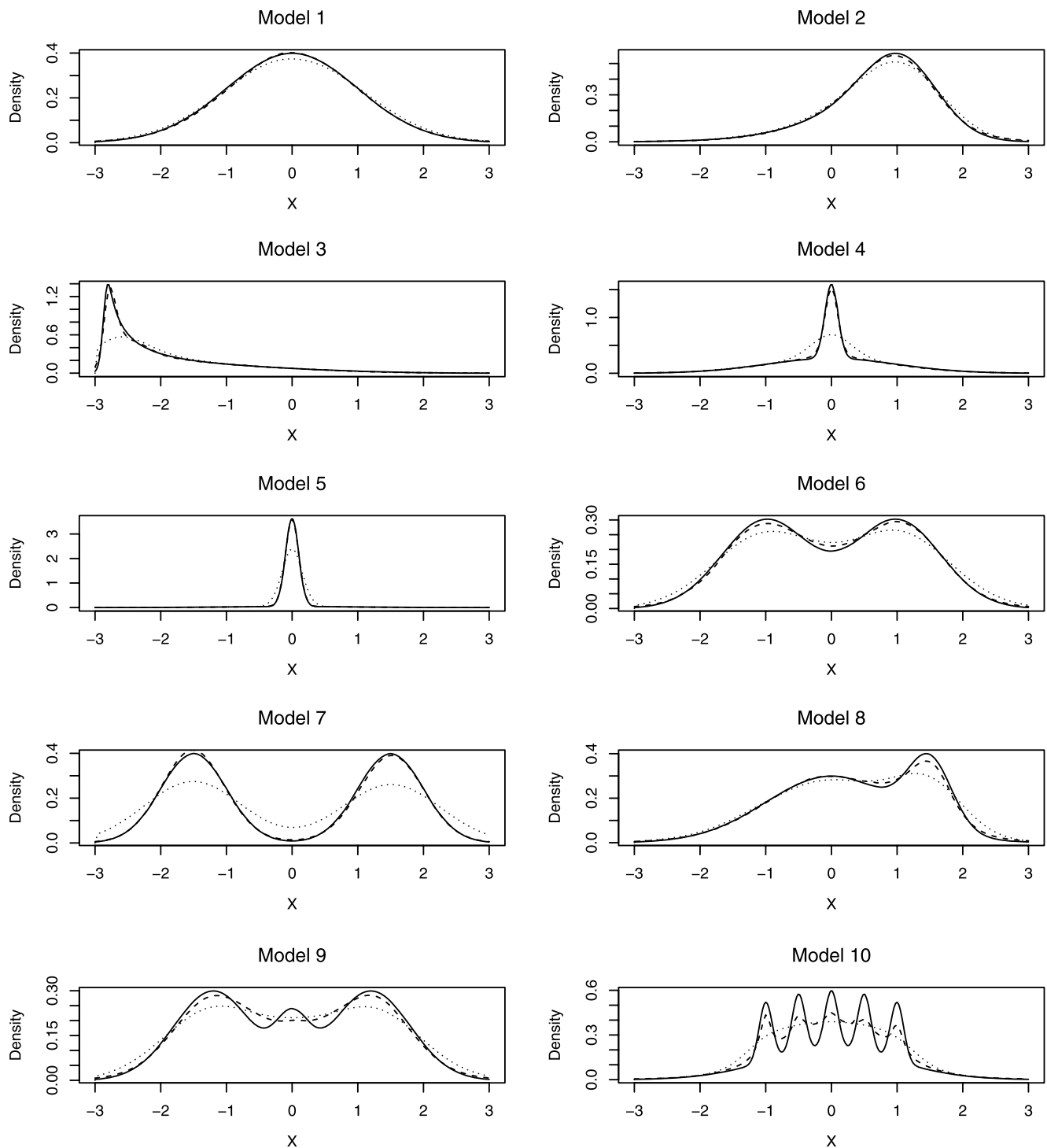}

\caption{Posterior density estimates for the first 10 models of Marron
and Wand (\citeyear{marron}): true density (solid line), NRMI normal mixture
estimate based
on the N-stable process (dashed line), and kernel density estimate
with optimal bandwith (dotted line). The estimates have been obtained
as averages over the $N=40$ simulation experiments.}\vspace*{-6pt}
\label{ex3sts}
\end{figure*}

Table~\ref{table3} summarizes the results in terms of RMISE. For
comparison purposes we have also included the RMISE obtained by \citet
{MV} using Bayesian wavelets and those obtained by \citet{roeder2} using
finite mixture of normals. In a private communication, M\"uller and
Vidakovic informed us of a minor problem with the RMISE values
originally reported in \citet{MV}: the values in Table~\ref{table3} are
the correct ones\vadjust{\goodbreak} obtained from their model. Figure~\ref{ex3sts}
displays the true density (solid line) and the estimated densities
resulting from our NRMI mixture (dashed line) and the kernel density
estimates with optimal bandwith (dotted line) for models 1--10. The
numbers reported in Table~\ref{table3} and the density estimates in
Figure~\ref{ex3sts} are averages over the 40 experiments.

From Table~\ref{table3} we can observe that the approach of \citet
{roeder2} improves on the kernel density estimator in $7$ of the $10$
models. In particular, they fail to provide a good fit for those
densities that are quite spiky (models $3$, $4$ and $10$). Also, the
wavelets approach of \citet{MV} have the best behavior precisely for
these spiky models producing the smallest RMISE.
The NRMI normal mixtures performs significantly better than the kernel
density estimator in all $10$ models, the highest RMISE being $0.86$.
This is also apparent in Figure~\ref{ex3sts}. Moreover, it reaches the
smallest RMISE in $6$ of the $10$ models compared to all its
competitors. However, rather than focusing on best performances, it is
important to stress that the estimates yielded by the approaches of R\&W and M\&V are, in some cases, significantly worse than the kernel
density estimator. Hence,\ NRMI mixtures give the best result in $6$
cases (models 3--5, $7$, $8$ and $10$), but, more importantly, yield
at least second-best results in all the other cases and there
is always quite some gap between its RMISE and the one of the worse
estimate. In summary, the flexibility of the NRMI mixtures makes it a
valuable alternative to more standard methods. In particular, the
N-stable mixtures could be considered as a default model, which works
reasonably well regardless of whether the density is unimodal,
multimodal, spiky or flat.

\begin{table}
\tabcolsep=0pt
\caption{RMISE statistic for the first $10$ models in Marron and Wand
(\citeyear{marron}): column two displays the RMISE values for the NRMI normal
mixture model based on a N-stable process; columns three and four
report the RMISE values for the methods of M{\"u}ller and Vidakovic (\citeyear{MV}) and Roeder and Wasserman
(\citeyear{roeder2}), respectively}\label{table3}
\begin{tabular*}{\columnwidth}{@{\extracolsep{4in minus 4in}}lccc@{}}
\hline
& \multicolumn{3}{c@{}}{\textbf{RMISE}} \\
 \ccline{2-4}
\textbf{Model}& \textbf{MRMI} & \textbf{M\&V} & \textbf{R\&W} \\
\hline
\phantom{0}1 & 0.39 & 1.99 & 0.07 \\
\phantom{0}2 & 0.76 & 0.98 & 0.34 \\
\phantom{0}3 & 0.18 & 0.28 & 2.91 \\
\phantom{0}4 & 0.09 & 0.25 & 1.67 \\
\phantom{0}5 & 0.05 & 0.43 & 0.44 \\
\phantom{0}6 & 0.81 & 1.62 & 0.31 \\
\phantom{0}7 & 0.13 & 0.38 & 0.23 \\
\phantom{0}8 & 0.73 & 1.72 & 0.74 \\
\phantom{0}9 & 0.86 & 1.42 & 0.54 \\
10 & 0.81 & 0.83 & 2.76 \\
\hline
\end{tabular*}
\end{table}

\begin{remark}
NRMI mixtures with nonparametric specification of both location
and scale parameters considered in this section correspond to the
\texttt{MixNRMI2} function in the {\textsf R}-package \texttt{BNPdensity}.
Additionally, the package also includes semi-param\-etric NRMI mixtures,
in which the location and the scale are modeled, respectively,
according to an\break NRMI and a parametric distribution.
Such a specification corresponds to a common value of the smoothing
parameter $\sigma$ for all mixture components and to locations $\mu
_j$'s generated by the NRMI. This is called the \texttt{MixNRMI1} function
in the package. Extensive simulation studies, not reported here,
indicate that semiparametric mixtures are more sensitive w.r.t. wrong
prior specifications, in the sense that they tend to get stuck on wrong
values for the number of mixture components. Moreover, as one would
expect given the lack of flexibility in controlling the dispersion,
some oversmoothing typically would appear.
\end{remark}

\begin{remark}
Although for comparison purposes it is more convenient to work
with simple NRMI mixtures as done here, extensions to more general
settings have been provided in the literature. For example,\vadjust{\goodbreak} \citet
{lijoi11} define vectors of dependent NRMIs, where the dependence
originates from a suitable construction of the underlying Poisson
random measures: such models are readily implementable in two-sample
problems and meta-analysis. More general regression problems can also
be obtained starting from simple NRMI mixtures. For instance, a
generalization of the ANOVA dependent Dirichlet process model \citep
{deiorio:04} to NRMI can be written via the hierarchical representation
\eqref{eq:mixt}. In the normal case the first
equation becomes
\[
X_i|\theta_i,Z_i,\sigma\simiid\no \bigl(
\theta_i'Z_i,\sigma^2 \bigr),
\]
where $Z_i$ is the covariate vector. The second and third equations
remain the same together with a prior specification for $\sigma$.
Suitable modifications of the simulation algorithm, and thus on the
\texttt{BNPdensi\-ty} package, can be implemented to cover this regression case.
\end{remark}

\section*{Acknowledgments} The authors are grateful to an Associate
Editor and three referees for insightful comments and suggestions. The
participants of the 2011 Bayesian nonparametrics research programme in
Veracruz, Mexico, where a preliminary version of this paper and the
corresponding {\textsf R}-package were presented, are also gratefully
acknowledged for several stimulating discussions. Special thanks are
due to Alejandro Jara, Peter M\"uller and Steven MacEachern for their
helpful suggestions which led to remarkable improvements of the {\textsf
R}-package.
E.~Barrios and L.~E. Nieto-Barajas supported by the National Council for Science and
Technology of Mexico (CONACYT), Grant I130991.
A. Lijoi and I.~Pr\"unster supported by the European Research Council (ERC) through
StG\break
\mbox{``N-BNP''} 306406.

%



\end{document}